\documentclass[%
 aip,
 amsmath,amssymb,
preprint,%
]{revtex4-1}

\usepackage{graphicx}% Include figure files
\usepackage{dcolumn}% Align table columns on decimal point
\usepackage{bm}% bold math
\usepackage[utf8]{inputenc}
\usepackage[T1]{fontenc}
\usepackage{mathptmx}
\usepackage{xcolor}

\begin{document}

%\preprint{AIP/123-QED}

%\title[Particle migration in blood flow]{A unified analysis of nano-to-microscale particle dispersion in tubular blood flow}
\title{A unified analysis of nano-to-microscale particle dispersion in tubular blood flow}
% Force line breaks with \\

\author{Z. Liu}
\email{zxliu@gatech.edu}
\affiliation{ 
The George W. Woodruff School of Mechanical Engineering, Georgia Institute of Technology, Atlanta, GA, 30332, USA %\\This line break forced with \textbackslash\textbackslash
}%
\affiliation{ 
The Parker H. Petit Institute for Bioengineering and Bioscience, Georgia Institute of Technology, Atlanta, GA, 30332, USA %\\This line break forced with \textbackslash\textbackslash
}%

\author{J. R. Clausen}%
%\email{jclause@sandia.gov}
\affiliation{Sandia National Laboratories, Albuquerque, NM, 87185, USA}%

\author{R. R. Rao}%
%\email{rrrao@sandia.gov}
\affiliation{Sandia National Laboratories, Albuquerque, NM, 87185, USA}%

% \author{D. N. Ku}
% %\altaffiliation[Also at ]{Parker H. Petit Institute for Bioengineering and Bioscience, Georgia Institute of Technology, Atlanta, GA, 30322, USA}
% %\homepage{http://www.Second.institution.edu/~Charlie.Author.}
% \email{david.ku@me.gatech.edu}
% \affiliation{ 
% The George W. Woodruff School of Mechanical Engineering, Georgia Institute of Technology, Atlanta, GA, 30332, USA %\\This line break forced with \textbackslash\textbackslash
% }%
% \affiliation{ 
% The Parker H. Petit Institute for Bioengineering and Bioscience, Georgia Institute of Technology, Atlanta, GA, 30332, USA %\\This line break forced with \textbackslash\textbackslash
% }%

\author{C. K. Aidun}
%\altaffiliation[Also at ]{Parker H. Petit Institute for Bioengineering and Bioscience, Georgia Institute of Technology, Atlanta, GA, 30322, USA}
%\homepage{http://www.Second.institution.edu/~Charlie.Author.}
\email{cyrus.aidun@me.gatech.edu \ (corresponding)}
\affiliation{ 
The George W. Woodruff School of Mechanical Engineering, Georgia Institute of Technology, Atlanta, GA, 30332, USA %\\This line break forced with \textbackslash\textbackslash
}%
\affiliation{ 
The Parker H. Petit Institute for Bioengineering and Bioscience, Georgia Institute of Technology, Atlanta, GA, 30332, USA %\\This line break forced with \textbackslash\textbackslash
}%

\date{\today}

\begin{abstract}
Transport of solid particles in blood flow exhibits qualitative differences in the transport mechanism when the particle varies from nanoscale to microscale size comparable to the red blood cell (RBC). 
The effect of microscale particle margination has been investigated by several groups. Also, the transport of nanoscale particles (NPs) in blood has received considerable attention in the past.
This study attempts to bridge the gap by quantitatively showing how the transport mechanism varies with particle size from nano- to microscale. Using a three-dimensional (3D) multiscale method, the dispersion of particles in microscale tubular flows is investigated for various hematocrits, vessel diameters and particle sizes. NPs exhibit a nonuniform, smoothly-dispersed distribution across the tube radius due to severe Brownian motion. The near-wall concentration of NPs can be moderately enhanced by increasing hematocrit and confinement. Moreover, there exists a critical particle size ($\sim$1 $\mu$m) that leads to excessive retention of particles in the cell-free region near the wall, i.e., margination. Above this threshold, the margination propensity increases with the particle size. The dominance of RBC-enhanced shear-induced diffusivity (RESID) over Brownian diffusivity (BD) results in 10 times higher radial diffusion rates in the RBC-laden region compared to that in the cell-free layer, correlated with the high margination propensity of microscale particles. This work captures the particle size-dependent transition from Brownian-motion dominant dispersion to margination using a unified 3D multiscale computational approach, and highlights the linkage between the radial distribution of RESID and the margination of particles in confined blood flows.
\end{abstract}

\maketitle

\section{Introduction}\label{sec:intro}
Blood is a complex fluid suspended with multiple species, primarily including red blood cells (RBCs), platelets, white blood cells and various biomolecules (such as von Willebrand factors, albumin, fibrinogen, etc.) that covers length scales ranging from nanometers to micrometers \citep{Perdikaris2016,Liu2018b}. In microvessels under physiological flow conditions, RBCs migrate towards the axis of the tube and leave a cell-free layer (CFL) near the wall~\citep{fung2013biomechanics,Secomb2017}. Such phenomenon, well known as the Fahraeus-Lindquist effect~\citep{fahraeus1931viscosity}, contributes to the hemorheological heterogeneity of the blood flow. Unraveling the dispersion properties of solutes and cells of various sizes ranging from nanometer to micrometer in such heterogeneous blood flows under vascular confinement can potentially lead to optimal design of drug carriers and better understanding, intervention and control of vascular diseases.

\textcolor{black}{As a relevant example of microscale particle transport in blood, platelets margination has shown to play an important role in affecting the rate of clot formation in hemostasis and thrombosis~\citep{Casa2017}. Motivated by that, a plethora of studies over the past decades have dedicated to unravel the mechanistic mechanisms of margination or segregation of microscale particles/cells in blood(-like) flows through perfusion experiements~\citep{Grabowski1972,aarts1988blood,eckstein1987,Ahmed2018}, continuum-level modeling~\citep{Eckstein1991,MM2015} and direct numerical simulations~\citep{fogelson2011jfm,Zhao2011,Zhao2012,Fedosov2012wbc,ReasorABE2013,MM2016,Ahmed2018,kruger2016,Bagchi2015sm,Gabor2019}. The platelet margination is found to be primarily driven by the cross-stream hydrodynamic fluctuation ~\citep{Zhao2011,Zhao2012,Shaqfeh2017PRF} or equivalently the RBC-enhanced shear-induced diffusion~\citep{MM2015,MM2016} in the RBC-laden region synergistically accompanied by the sink-like effect of the CFL~\citep{MM2016}.}

Nanoscale particle (NP) dispersion in blood flow, on the other end of the spectrum, has recently received considerable attention due to the fast development of nano-drug delivery techniques that have the potential to revolutionize the traditional therapeutics~\citep{Albanese2012}. Although the effective diffusivity of nanoscale solutes in blood flow were measured decades ago~\citep{Diller1980}, it is not until the past several years multiscale particle-level simulation techniques~\citep{Tan2011,Lee2013,Muller2014,Liu2018a} become feasible. Tan $et\ al.$~\citep{Tan2011} apply a coupled Brownian dynamics and immersed finite-element (FE) method to study the influence of RBCs on the NP dispersion in blood flows, showing substantial margination behavior for 100 nm particles. Through both in vivo and in silico techniques, \citet{Lee2013} show that submicron particles (>500 nm) can marginate while NPs ($\sim$100 nm) are mostly trapped in the RBC-laden region. Muller $et\ al.$ \citep{Muller2014} performed two-dimensional (2D) simulations and suggest that microscale particles compared to submicroscale particles show better margination propensity. \citet{Liu2018a} develop a multiscale complex blood solver and evaluate the role of BD versus RESID in affecting the biodistribution of NPs. Recently, \citet{Liu2019} characterize the complete 3-D diffusivity tensor of NP in blood under various shear rates and hematocrits, which can be employed to modeling large-scale NP biotransport applications. 

Although the transport of both nanoscale and microscale particles in blood have been understood to a large extent, there is still a lack of a systematic interrogation of the particle dispersion behavior across nano-to-microscale sizes using a unified computational approach. Consequently, questions such as whether nanoscale particles exhibit margination qualitatively the same as microscale particle does still remains controversial. A recent effort by~\citet{cooley2018influence} using in vitro experiment and 2D in silico simulation to understand the cross-length-scale particle margination and adhesion propensity has set an example for a unified understanding of the nano-to-microscale particle dispersion in blood flows. However, the general physical mechanisms behind the multiscale particle dispersion/margination phenomenon in blood are still not presented; besides, the 2D simulation could still overlook the 3D nature of the tubular blood flow phenomena.   

In this work, we employ a recently developed 3D multiscale and multicomponent blood flow solver~\citep{Reasor2012,Liu2018a,Liu2018b,Liu2019} to tackle the dispersive characteristics of spherical, rigid particles with sizes spanning nano-to-microscale in a tubular blood flow. Particle suspension dynamics in the presence of thermal fluctuation, RBC-particle direct and hydrodynamic interactions and wall-bounded confinement effect are captured under a unified 3D computational framework. The strong correlation between the non-uniform distribution of particle radial diffusivity and the equilibrium distribution of particle radial concentration is highlighted to gain mechanistic understanding of the occurrence of particle-size-induced dispersion-to-margination transition.

The remainder of the paper is organized as follows. In \S \ref{sec:method}, we describe the unified multiscale complex blood solver and layout the techniques for evaluation of particle radial concentration and diffusivity. In \S \ref{sec:res}, we present the simulation results, where the particle radial distribution at equilibrium state is discussed under various confinement ratio, hematocrit and particle sizes. The mechanisms that drives the particle size-dependent dispersion-to-margination transition will be discussed. In \S \ref{sec:conclusion}, we conclude this systematic study.

%%%%%%%%%%%%%%%%%%%%%%%%%%%%%%%%%%%%%%%%%%%%%%%%%%%%%%%%%%%%%%%%%%%%%%%%%%%%%%
%%%%%%%                           Method                                %%%%%%
%%%%%%%%%%%%%%%%%%%%%%%%%%%%%%%%%%%%%%%%%%%%%%%%%%%%%%%%%%%%%%%%%%%%%%%%%%%%%%
\section{Methodology}\label{sec:method}
\begin{figure}
\begin{center}
\includegraphics[width = 0.8\textwidth]{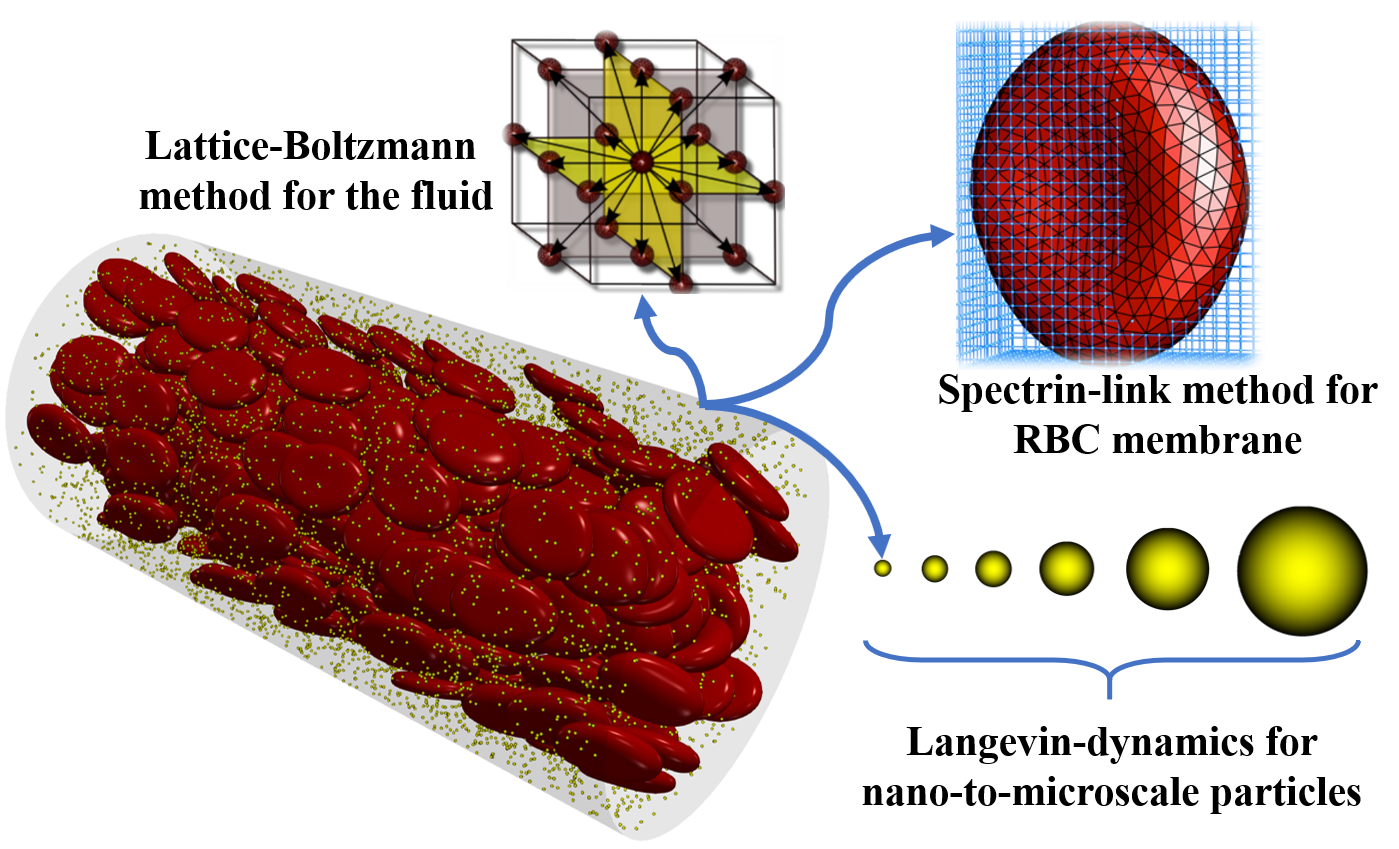}
\caption{Nano-to-microscale particle transport in cellular blood flow through microvessels. The fluid phase is simulated using the lattice-Boltzmann (LB) method~\citep{Aidun2010}. The deformation and dynamics of red blood cells (RBCs) are simulated by coupling a course-grained spectrin-link (SL) method with LB method~\citep{Reasor2012}. The multiscale (nanoscale to microscale) particles (yellow) are simulated via a coupled LB-Langevin dynamics (LD) method~\citep{Liu2018a,Liu2018b}. The particle-RBC interaction and inter-cell interactions are resolved through various contact modeling techniques~\citep{MacJFM2009,ClausenJFM2011,Liu2018a,Liu2019}.}
\label{fig:frame}
\end{center}
\end{figure}

The numerical method used to simulate the bi-disperse particle-RBC suspensions confined in a tubular flow is through a multiscale and multicomponent complex blood flow method \citep{Liu2018a,Liu2019} that couples the lattice-Boltzmann/Spectrin-link (LB-SL) method \citep{Reasor2012} with the lattice-Boltzmann/Langevin-dynamics (LB-LD) method \citep{Liu2018b}. This method leverages the off-lattice nature of the LB-LD approach and the efficiency of the course-grained SL RBC membrane method to concurrently simulate the dynamics of across nano-to-microscale particles and microscale deformable capsules with a fixed LB lattice resolution \citep{Liu2018a,Liu2018b}. The hybrid LB-LD-SL method has previously been verified with theory \citep{Liu2018a,Liu2018b} and validated against experiments \citep{Reasor2012,ReasorJFM2013,Liu2019}. Fig~\ref{fig:frame} demonstrates a nanoscale particle-RBC bidisperse suspension flow through a 40 $\mu m$ vessel, where the computational methods for each module are denoted accordingly and presented in detail as follows.

\subsection{\label{sec:lbm} Lattice-Boltzmann method}
Simulation of the suspending fluid is based on the Aidun-Lu-Ding (ALD) LB method\citep{Aidun1995,Aidun1998,Aidun2010}. The LB method solves the discretized Boltzmann transport equation in velocity space through the streaming-collision process. In streaming, the fictitious fluid particles propagate along discrete velocity vectors forming a lattice space. In collision, the fluid particles at each lattice site collide with each other, causing the relaxation of the particle distribution function (PDF) towards a local `Maxwellian' equilibrium PDF. The collision term is linearized based on the single-relaxation-time Bhatnagar, Gross, and Krook (BGK) operator \cite{Bhatnagarp}. The temporal evolution of the particle distribution function is given as
\begin{equation}
\begin{aligned}
  f_i(\boldsymbol{r}+\Delta t \boldsymbol{e}_i, t+\Delta t) =  f_i(\boldsymbol{r}, t) - \frac{\Delta t}{\tau}[f_i(\boldsymbol{r}, t) - f_i^{(0)}(\boldsymbol{r}, t)] + f_i^S(\boldsymbol{r}, t),
  \label{eqn:lb1}
\end{aligned}
\end{equation}
where $f_i$ is the fluid PDF, $f_i^{(0)}$ is the equilibrium PDF, $r$ is the lattice site, $e_i$ is the discrete lattice velocity, $t$ is time, $\tau$ is the single relaxation time and $f_i^S$ is a forcing source term introduced to account for the discrete external force effect. The method has a pseudo speed of sound, $c_s=\Delta r/(\sqrt{3} \Delta t)$, and a fluid kinematic viscosity, $\nu$=$(\tau - \Delta t/2) c_s^2$, where $\Delta t$ is the time step and $\Delta r$ is the unit lattice distance. The positivity of $\nu$ requires $\tau$$>$$\Delta t/2$. In the LB method, time and space are typically normalized by $\Delta t$ and $\Delta r$, respectively, such that $\Delta t_{LB}$=$\Delta r_{LB}$=1 are employed to advance equation \ref{eqn:lb1}. In the near incompressible limit (i.e., the Mach number, $Ma$=$u/c_s$$\ll$1), the LB equation recovers the Navier-Stokes equation \cite{Junk2003} with the equilibrium PDF given in terms of local macroscopic variables as
\begin{equation}
\begin{aligned}
  f_i^{(0)}(\boldsymbol{r}, t) = \omega_i \rho [1 +  \frac{1}{c_s^2}(\boldsymbol{e}_i \cdot \boldsymbol{u}) + \frac{1}{2c_s^4}(\boldsymbol{e}_i \cdot \boldsymbol{u})^2 - \frac{1}{2c_s^2}(\boldsymbol{u} \cdot \boldsymbol{u})],
\end{aligned}
  \label{eqn:lb2}
\end{equation}
where $\omega_i$ denotes the set of lattice weights defined by the LB stencil in use. The macroscopic properties such as the fluid density, $\rho$, velocity, $\boldsymbol{u}$, and pressure, $p$, are obtained via moments of the equilibrium distribution functions as, $\rho=\sum_{i=1}^Q f_i^{(0)}(\boldsymbol{r},t)$, $\boldsymbol{u}=\frac{1}{\rho}\sum_{i=1}^Q f_i^{(0)}(\boldsymbol{r},t) \boldsymbol{e}_i$ and $p\mathbf{I}=\sum_{i=1}^Q f_i^{(0)}(\boldsymbol{r},t) \boldsymbol{e}_i\boldsymbol{e}_i-\rho\boldsymbol{u}\boldsymbol{u}$, respectively. Here, $\mathbf{I}$ is the identity tensor and pressure can be related to density and the speed of sound through $p$=$\rho c_s^2$. For the D3Q19 stencil adopted in the current study, $Q$ is equal to 19. Along the rest, non-diagonal, and diagonal lattice directions, $\omega_i$ is equal to 1/3, 1/18, and 1/36, and $|\boldsymbol{e}_i|$ is equal to 0, $\Delta r/\Delta t$, and $\sqrt{2}(\Delta r/\Delta t)$, correspondingly.

\subsection{Langevin-dynamics method}\label{sec:ld}
The nano-to-microscale particle suspensions are resolved through a two-way coupled LB-LD method which has been verified~\citep{Liu2018a,Liu2018b} and validated against experiments~\citep{Liu2019}. This approach treats suspended particles in Stokesian regimes as point particles, while the volume exclusion effect of the particles are resolved through potential equations. The dynamics of LD particles is governed by the Langevin equation (LE),
\begin{equation}
  m_p \frac{d \boldsymbol{u}_p}{dt} = \boldsymbol{C}_p + \boldsymbol{F}_p + \boldsymbol{S}_p,
  \label{eqn:ld1}
\end{equation}
where $m_p$ is the mass of a single particle. The conservative force, $\boldsymbol{C}_p$, specifying the interparticle and particle-surface interaction forces, is determined by calculating the directional derivatives of the total potential energy $U_{total}$ as
\begin{equation}
  \boldsymbol{C}_p = - \frac{dU_{total}}{d\boldsymbol{r}_p},
  \label{eqn:ld2}
\end{equation}
where in this study $U_{total}$ accounts for the particle-cell/wall short-distance interactions, as discussed in \S \ref{sec:contact}. The frictional force, $\boldsymbol{F}_p$, is assumed to be proportional to the relative velocity of the particle with respect to the local viscous fluid velocity \cite{Ahlrichs1998,Ahlrichs1999},
\begin{equation}
  \boldsymbol{F}_p = -\zeta[\boldsymbol{u}_p(t)-\boldsymbol{u}(\boldsymbol{r}_p,t)],
  \label{eqn:ld3}
\end{equation}
where $\boldsymbol{u}_p$ denotes the particle velocity, and $\boldsymbol{u}(\boldsymbol{r}_p,t)$ is the interpolated LB fluid velocity at the center of the particle. The friction coefficient, $\zeta$, is determined by the Stokes’ drag law, $\zeta=3\pi\mu d_p$, where $\mu$ is the dynamic viscosity of the suspending fluid. The stochastic force, $\boldsymbol{S}_p$, explicitly gives rise to the Brownian motion of the particle and satisfies the fluctuation-dissipation theorem (FDT) \cite{Kubo1966} by
\begin{equation}
  \langle S_{p,i}^{\alpha} (t)\rangle = 0, \ \ 
  \langle S_{p,i}^{\alpha} (t)S_{p,j}^{\beta} (t)\rangle = 2k_BT\zeta \delta_{ij}\delta_{\alpha\beta}\delta(t-t'),
  %\label{eqn:ld4}
  %\langle S_{p,i}^{\alpha} (t)S_{p,j}^{\beta} (t)\rangle = 2k_BT\zeta \delta_{ij}\delta_{\alpha\beta}\delta(t-t'),
  \label{eqn:ld5}
\end{equation}
where $i,j\in \{ x,y,z\}$, $\alpha$ and $\beta$ run through all the particle indices, $\delta_{ij}$ and $\delta_{\alpha\beta}$ are Kronecker deltas, $\delta(t-t')$ is the Dirac-delta function, $k_B$ is the Boltzmann constant and $T$ is the absolute temperature of the suspending fluid. The angle brackets denote the ensemble average over all the realizations of the random variables. Since we are concerned with long-time scale phenomenon, the over-damped LE is adopted in the current study as suggested in \citet{Liu2018a,Liu2018b}. 

\subsection{Spectrin-link method}\label{sec:sl}
The modeling of RBC dynamics and deformation is through the coarse-grained spectrin-link (SL) membrane method \cite{Pivkin2008,FedosovBJ2010} coupled to the LB method  \cite{Reasor2012}. The hybrid LB-SL method has been extensively validated against experimental measurements and is capable of capturing both the deformation and dynamics of single RBC \cite{Reasor2012} and the rheology of RBC suspensions at physiological hematocrit \cite{ReasorJFM2013} with good accuracy and efficiency. 

In the LB-SL model, the RBC membrane is modeled as a triangulated network with a collection of vertices mimicking actin vertex coordinates. The Helmholtz free energy of the network system, $E$, including in-plane, bending, volume and surface area energy components \cite{Dao2006}, is given by 
\begin{equation}
  E = 
  E_{IP} + E_{B} + E_{\Omega} + E_{A},
  \label{eqn:sl1}
\end{equation}
where the in-plane energy, $E_{IP}$, characterizes the membrane shear modulus through a worm-like chain (WLC) potential \cite{Bustamante2003} coupled with a hydrostatic component \cite{FedosovBJ2010}; the bending energy, $E_B$, specifies the membrane bending stiffness, which is essential in characterizing the equilibrium RBC biconcave morphology \cite{Dao2006,FedosovBJ2010}; the volumetric contraint energy, $E_{\Omega}$, and the area constraint energy, $E_A$, preserve the RBC volume and area conservation, respectively, when subject to external forces. 

The dynamics of each vertice are updated according to the Newton$\text{'}$s equations of motion,
\begin{equation}
  \frac{d\boldsymbol{x}_n}{dt}=\boldsymbol{v}_n,\ \ M\frac{d\boldsymbol{v}_n}{dt}=\bold{f}_n^{SL}+\bold{f}_n^{LB}+\bold{f}_n^{CC}
  \label{eqn:sl2}
\end{equation}
where $\boldsymbol{v}_n$ is the velocity of the vertice at the position $x_n$ and \textcolor{black}{$M$ is taken as the fictitious mass of the RBC that is evaluated as the total mass of the cell divided by the number of vertices, $N_v$ \citep{ReasorABE2013,Liu2019}.} The number of vertices used to discretize the RBC membrane is $N_v$=613, which has shown to yield adequate resolution to resolve the hydrodynamic forces \cite{MacJFM2009} and capture single RBC dynamics \cite{Reasor2012} and concentrated RBC suspension rheology \cite{ReasorJFM2013} when coupled with the LB method. $\bold{f}_n^{LB}$ specifies the forces on the vertex due to the fluid-solid coupling. $\bold{f}_n^{CC}$ are the forces due to cell-cell interactions. The forces due to the Helmholtz free energy based on the SL model is determined by
\begin{equation}
  \bold{f}_n^{SL} = -\frac{\partial E(\boldsymbol{x}_n)}{\partial \boldsymbol{x}_n}.
  \label{eqn:sl3}
\end{equation}
The SL method is solved by integrating equations \ref{eqn:sl2} at each LB time step using a first-order-accurate forward Euler scheme in consistency with the LB evolution equation to avoid excessive computational expense \cite{Reasor2012,Liu2018a}.

\subsection{\label{sec:fsi} Fluid-RBC coupling}
The coupling between fluid and RBC is accomplished through the ALD fluid-solid interaction scheme \citep{Aidun1998}. In this method, the momentum transfer at the fluid-solid interface is accounted for by applying the bounce-back operation along lattice links that cross solid surfaces. As a result, the no-slip condition is enforced by adjusting the PDFs of the fluid nodes at the end point of a link along the $i$ direction through
\begin{equation}
  f_{i'}(\boldsymbol{r},t+1) = f_i(\boldsymbol{r},t^+)-6\rho \omega_i \boldsymbol{u}_b\cdot \boldsymbol{e}_i,
  \label{eqn:fsi1}
\end{equation}
where $i'$ is the direction opposite to $i$, $f_i (\boldsymbol{r},t^+)$ is the post-collision distribution, and $\boldsymbol{u}_b$ is the solid velocity at the intersection point with the link. The fluid force exerted on the vertex on the RBC membrane mesh can be determined by
\begin{equation}
  \mathbf{f}_n^{LB}(\boldsymbol{r}+\frac{1}{2}\boldsymbol{e}_i,t) = 2\boldsymbol{e}_i [f_i(\boldsymbol{r},t^+)+3\rho \omega_i \boldsymbol{u}_b\cdot \boldsymbol{e}_{i'}],
  \label{eqn:fsi2}
\end{equation}
which is applied to the advancement of the RBC dynamic equation through equation~\ref{eqn:sl2}.

\subsection{Fluid-particle coupling}
\textcolor{black}{The LD particles with Brownian effect are coupled to the non-fluctuating LB fluid in a two-way fashion using spatial extra/inter-polation schemes \cite{Ahlrichs1999,Peskin2002}, through which treatment the long-distance many-body hydrodynamic interactions and the correct temperature scale can be captured simultaneously without empirical re-normalization \cite{Mynam2011,Liu2018a,Liu2018b}.} Specifically, the hydrodynamic force exerted on the particle, $\boldsymbol{F}_p^H$, is systematically decomposed into frictional and stochastic components as
\begin{equation}
  \boldsymbol{F}_p^H = \boldsymbol{F}_p+\boldsymbol{S}_p=-\zeta[\boldsymbol{u}_p(t)-\boldsymbol{u}(\boldsymbol{r}_p,t)]+\boldsymbol{S}_p,
  \label{eqn:fsi3}
\end{equation}
where the fluid velocity at the particle site, $\boldsymbol{u}(\boldsymbol{r}_p,t)$, is interpolated based on surrounding LB velocities and applied to update the LD particle dynamics through equation \ref{eqn:ld1}. The weighting functions, $w(\boldsymbol{r},\boldsymbol{r}_p)$, for interpolation is constructed using a trilinear stencil \cite{Ahlrichs1998,Liu2018a}. Since $\boldsymbol{F}_p$ and $\boldsymbol{S}_p$ are both originated from the `collision' between NP and liquid molecules, $\boldsymbol{F}_p^H$ (instead of $\boldsymbol{F}_p$) is assigned back to the fluid phase to satisfy momentum conservation. The same weighting function is then applied to constructing the local forcing source term as
\begin{equation}
  f_{i}^S(\boldsymbol{r},t) = - \frac{w(\boldsymbol{r},\boldsymbol{r}_p) \omega_i \boldsymbol{F}_p^H \cdot \boldsymbol{e}_i}{c^2_s\Delta r^3},
  \label{eqn:fsi4}
\end{equation}
which is adopted by equation \ref{eqn:lb1} to update the local hydrodynamics. The coupled LB-LD method, similar to the external boundary force (EBF) method \cite{Wu2008}, modifies the conventional LB evolution equation into equation \ref{eqn:lb1} by adding the forcing distribution function $f_i^S(\boldsymbol{r},t)$, which is shown to approximate the Navier-Stokes equation in the macroscopic scale ~\citep{guo2002}.

\begin{figure}[h]
	\centering
	\includegraphics[width=0.8\linewidth]{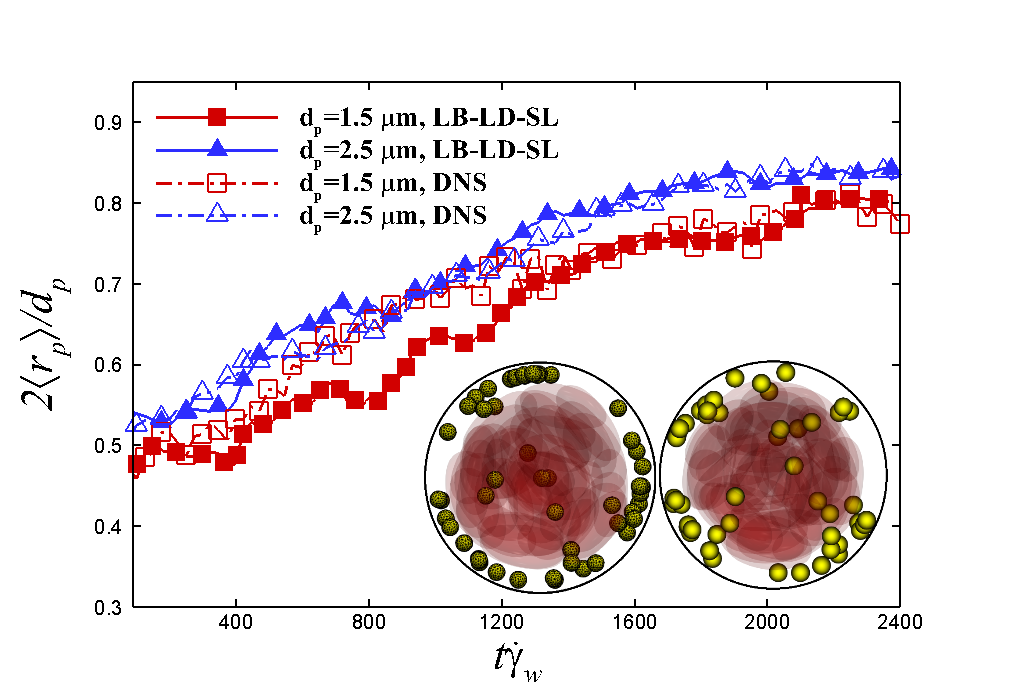}
\caption{\footnotesize \textcolor{black}{Time change of the average radial location of the microscale particles simulated using the multiscale LB-LD-SL method~\citep{Liu2018a,Liu2019} and the DNS approach~\citep{ReasorABE2013}. The tube diameter is 20 $\mu m$; the wall shear rate is 1000 $s^{-1}$; the hematocrit is 20\%. Particles with diameters of $d_p$=1.5 or 2.5 $\mu m$ have been selected for comparisons. The inset shows snapshots (frontal views) of 1.5 $\mu m$ particle distribution in tubular blood flows at $t\dot{\gamma}_w$=2000 simulated using the DNS method~\citep{ReasorABE2013} (left) or the LB-LD-SL multiscale approach~\citep{Liu2018a,Liu2019} (right).}}
	\label{fig:cnt}
\end{figure}

\subsection{\label{sec:contact} Contact modeling}
The short-distance interactions between particle and RBC or between particle and the vessel wall is through Morse potential that forbids particles from penetrating the RBC membrane or the vascular wall. This contact model has previously been used in the characterization of the NP long-time diffusion tensor in an unbounded sheared blood, where the calculated NP diffusivity compares favorably with experimental measurements \citep{Liu2019}. The Morse potential function is given as
\begin{equation}
  U_{M}(r)=D_e[e^{-2\beta(r-r_0)}-2e^{-\beta(r-r_0)}],\ \ (r\leq r_0) 
  \label{eqn:Morse}
\end{equation}
where $r$ is the normal distance between the particle center to the RBC surface, $r_0$ is a cut-off distance in which no interaction forces are present, $D_e$ is the potential well depth and $\beta$ is a scaling factor. The Morse potential is imposed when $r\leq r_0$ to preserve the repulsive effect. Model parameters are adjusted to match the measured inter-cell potential energy, as discussed in \citet{Liu2018a,Liu2019}. 
\textcolor{black}{Specifically, the scaling factor is set to $\beta=2\ \mu m^{-1}$, the surface energy has a value of $D_e=10^7k_BT$ and the equilibrium distance is set to $r_0=d_p/2+10\ nm$.
This simple contact model, bridging the LB-LD approach \citep{Liu2018a,Liu2018b} and the LB-SL method \citep{Reasor2012,ReasorJFM2013}, can capture the margination phenomenon of microscale particles comparably well as the DNS approach does~\citep{Reasor2012,MM2016}. Fig~\ref{fig:cnt} presents the temporal evolution of the ensemble average of the radial displacement of microscale particles, $2\langle r_p\rangle/d_p$, where the particle margination process through the LB-LD-SL approach and that via DNS compares favorably well especially when approaching the equilibrium stage ($t\dot{\gamma}_w\ge$2000).}

\subsection{\label{sec:radcon} Evaluation of the particle radial concentration}
The particle number concentration at specific radial location, $C_n(r,t)$, can be evaluated as
\begin{equation}
C_n(r,t) = \frac{\sum_{\alpha\in N}\{\delta[r_p^\alpha(t)-r]\}}{2\pi r\Delta rL_v},
\label{eqn:radcon}
\end{equation}
where $N$ denotes all LD particles in the simulation and $L_v$ is the length of the tube. The radial bin width, $\Delta r$, is set to one tenth of the tube radius to accurately resolve the radial profiles of the particle concentration distribution\citep{ReasorABE2013}. The bulk ensemble-averaged particle number concentration can be calculated as $\langle C_n\rangle$=$4N/\pi d_v^2L_v$, which is later used to normalize the pariticle local concentration.

\subsection{\label{sec:raddiff} Evaluation of the particle radial diffusivity}
The particle radial diffusivity is evaluated through a moving time-origin measurement \citep{Bolintineanu2014} of the particle mean squared displacement (MSD) based on a fixed sampling time interval (STI). The STI is properly chosen to exclude the short-time ballistic regime \citep{Liu2018a,Liu2018b}. By measuring the radial MSD of particles at a radial location $r$, the local instantaneous particle radial diffusivity can be evaluated according to
\begin{equation}
D_{rr}(r,t) = \frac{\sum_{\alpha\in N}\{\delta[r_p^\alpha(t)-r][r_p^\alpha(t+\Delta t)-r_p^\alpha(t)]^2\}}{2\Delta t\sum_{\alpha\in N}\{\delta[r_p^\alpha(t)-r]\}},
\label{eqn:raddiff}
\end{equation}
where $N$ denotes all LD particles in the simulation and $\Delta t$ is chosen to be $1000$ in lattice units \citep{Liu2019}. Same technique can be applied to measure the radial distribution of RESID, $D_{rr}^{RBC}(r,t)$,  where the BD is excluded by setting $\boldsymbol{S}_p=0$. The bulk ensemble-averaged particle radial diffusivity is calculated as $\langle D_{rr}(t)\rangle$=$\frac{\sum_{\alpha\in N}\{[r_p^\alpha(t+\Delta t)-r_p^\alpha(t)]^2\}}{2\Delta t}$; similarly, the bulk ensemble-averaged RESID can be obtained as $\langle D_{rr}^{RBC}(t)\rangle$=$\frac{\sum_{\alpha\in N}\{[r_p^\alpha(t+\Delta t)-r_p^\alpha(t)]^2\}}{2\Delta t}|_{\boldsymbol{S}_p=0}$. The equilibrium counterparts of the particle radial diffusivity are denoted as $\langle D_{rr}\rangle$ and $\langle D_{rr}^{RBC}\rangle$ without time dependence.

%%%%%%%%%%%%%%%%%%%%%%%%%%%%%%%%%%%%%%%%%%%%%%%%%%%%%%%%%%%%%%%%%%%%%%%%%%%%%%
%%%%%%%                           Results                               %%%%%%
%%%%%%%%%%%%%%%%%%%%%%%%%%%%%%%%%%%%%%%%%%%%%%%%%%%%%%%%%%%%%%%%%%%%%%%%%%%%%%
\section{Simulation results}\label{sec:res}
\subsection{Setup}\label{sec:st}
The physical problem of particle-RBC suspension flow through a straight tube can be defined by the vessel diameter, $d_v$, the systemic hematocrit, $\phi$, the particle diameter, $d_p$, the wall shear rate, $\dot{\gamma}_w$, and temperature, $T$, given fixed RBC properties (hydrodynamic radius, $a_{RBC}$, and membrane shear modulus, $G$). Apart from the hematocrit, the corresponding non-dimensional parameters are the confinement ratio, $d_v^* = \frac{a_{RBC}}{d_v}$, which determines the severity of the RBC finite size effect; the particle-cell size ratio, $d_p^* = \frac{d_p}{a_{RBC}}$, that quantifies the length-scale discrepancy between the two species suspended; the Peclet number, $Pe = \frac{3\mu\pi\dot{\gamma}_wd_pa_{RBC}^2}{k_BT}$, which describes the competition between the shear-induced diffusion and the Brownian diffusion; and the capillary number, $Ca = \frac{\mu\dot{\gamma}_wa_{RBC}}{G}$, which defines the deformability of the RBC capsule.

\begin{figure}[h]
\centering
\includegraphics[width=0.68\linewidth]{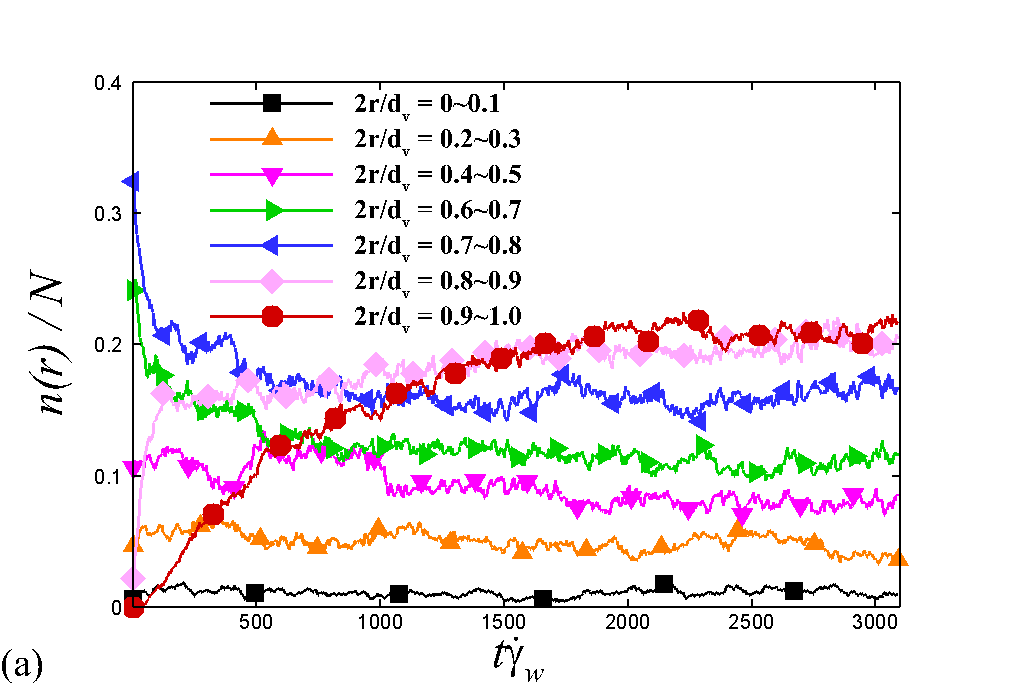}
\includegraphics[width=0.68\linewidth]{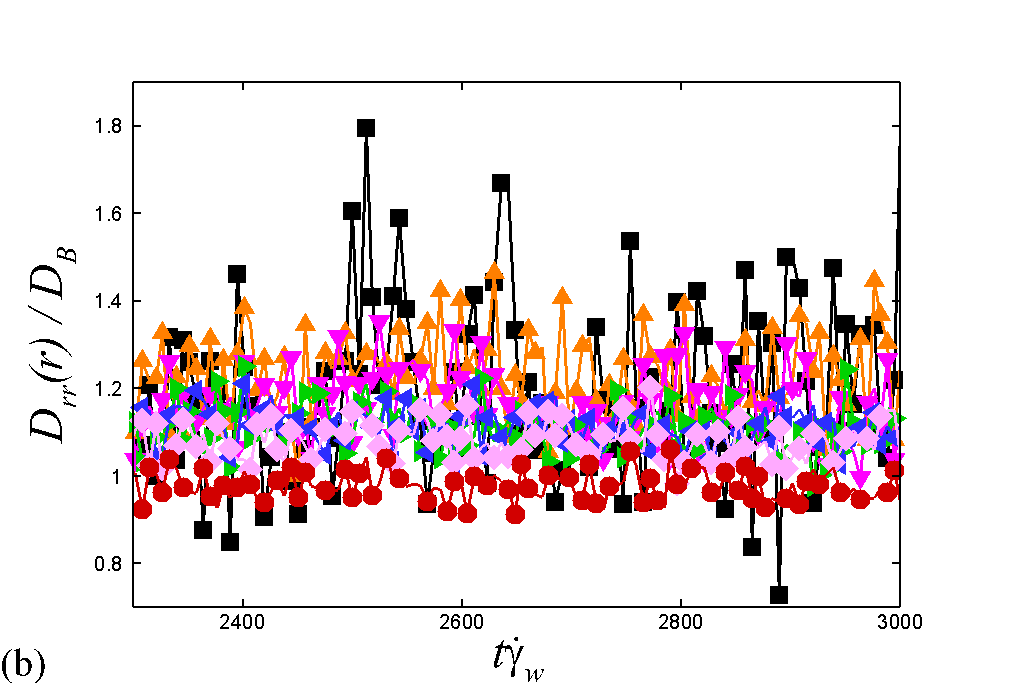}
\caption{
\footnotesize Temporal change of (a) particle number percentage and (b) particle radial diffusivity at different radial locations. Here, the particle number percentage, $n(r)/N$, is defined as the number of particles within certain peripheral layer, $n(r)$, normalized by the total particle number, $N$. Simulation is performed with $d_p$=100 $nm$, $d_v$=20 $\mu m$ and $\dot{\gamma}_w$=1000 $s^{-1}$. Simulation reaches equilibrium after $t\dot{\gamma}_w\sim2000$.}
\label{fig:evl}
\end{figure}

In this work, we consider $d_v^*$ ranging from 0.07$\sim$0.29, corresponding to typical diameters of arterioles~\citep{Lipo2005}. The particle-cell size ratio considered ranges from $d_p^*$=0.003$\sim$0.86, covering typical size of biomolecules and cells (such as von Willebrand factor, vWF, and platelet) in blood flows. Given the low sensitivity of platelet margination to shear rate~\citep{MM2016}, a physiologically relevant wall shear rate, $\dot{\gamma}_w$=1000 $s^{-1}$, typical in arterioles or capillaries is considered for all cases. The fluid viscosity is set to the same as blood plasma, $\mu$= 1.2 $cp$. The temperature is set to the body temperature, $T$= 310 K. The RBC membrane has a shear modulus of $G$=0.0063 $dynes/cm$. The effective hydrodynamic radius of RBC is $a_{RBC}$= 2.9 $\mu m$. As a result, the dependence on $Pe$ is determined by $d_p^*$. The deformability of RBC is fixed with $Ca_G$=0.55.

\textcolor{black}{All simulations are initialized with the particles and RBCs uniformly and randomly mixed in the tube, except the particles are only seeded at $2r/d_v\leq0.6$. Periodic boundary conditions are imposed on the two ends of the tube. The tube has a length of $L_v/a_{RBC}\ge10$ to ensure the periodic boundary treatment exerting negligible effect on the particle/cell transport. This is paper focuses on the dispersive characteristics at equilibrium, albeit the transient effects may play a significant role in the particle distribution in microvascular bifurcating structures\citep{iori2015,bacher2018,Bagchi2018POF,Liu2018bif,balabani2018}. The equilibrium conditions are determined by tracking the particle accumulation at each radial location until it plateaus. As an example in Fig \ref{fig:evl}, we present the temporal change of particle number percentage and radial diffusivity at different radial locations; where the particle number percentage, $n(r)/N$, is defined as the number of particles within certain radial layer, $n(r)=\sum_{\alpha\in N}\{\delta[r_p^\alpha(t)-r]\}$, normalized by the total particle number, $N$, within the simulation domain. The simulation is performed with $d_p$=100 $nm$, $d_v$=20 $\mu m$ and $\phi$=0.2. The equilibrium state is arrived at $t\dot{\gamma}_w\sim2000$, when the mean values of both $n(r)/N$ and $D_{rr}(r)/D_B$ remain unchanged with respect to time.}

\subsection{Dependence on confinement}\label{sec:conf}
We first interrogate the dispersion characteristics of NPs under different confinement ratios controlled by adjusting vessel diameters in the range of $d_v$=10$\sim$40 $\mu m$ (corresponding to $d_v^*$=0.29$\sim$0.073), which corresponds to typical size of arterioles or capillaries in human \citep{Lipo2005}. The particle size is fixed to $d_p$=100 $nm$. The wall shear rate is set to $\dot{\gamma}_w$=1000 $s^{-1}$ and the systemic hematocrit is set to $\phi$=0.2, which are within the range of physiological hemorheological ranges in human arterioles or capillaries \citep{Lipo2005}. The number of particles simulated in the microvessels are $N$=4000, 1000 and 250 from large to small vessels, respectively, to conserve the particle volume concentration.
\begin{figure}[h]
\centering
\includegraphics[width=0.96\linewidth]{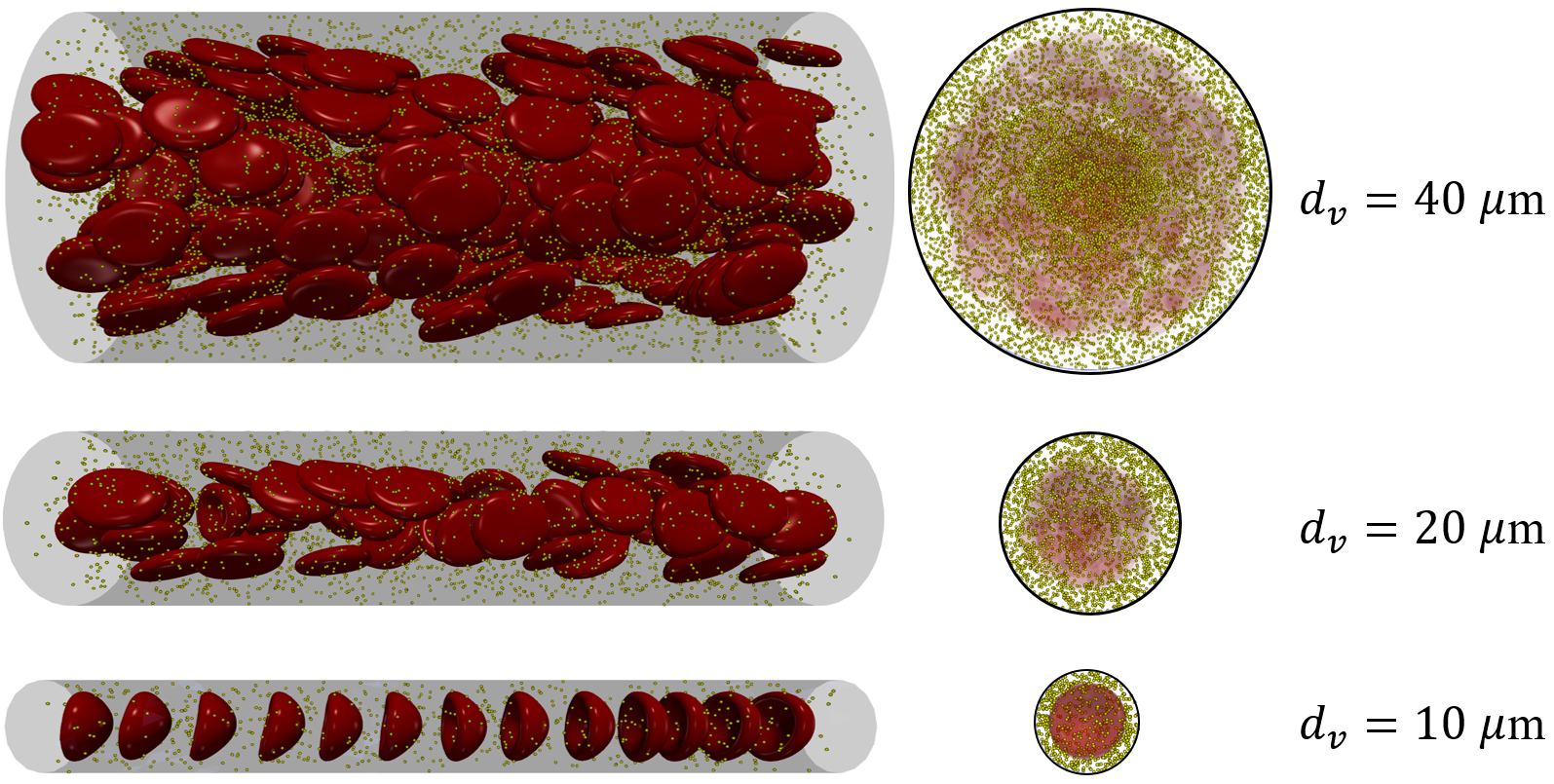}
\caption{\footnotesize NP and RBC distribution at equilibrium within microvessels of different diameters $d_v=40$ (top), $20$ (middle) or $10$ (bottom) $\mu m$ at $\phi=0.2$ and $\dot{\gamma}_w=1000\ s^{-1}$. Left columns show the side views of the microvessels; right columns show the end views of the microvessels.}
\label{fig:conf1}
\end{figure}

\begin{figure}[h]
\centering
 \includegraphics[width=0.47\linewidth]{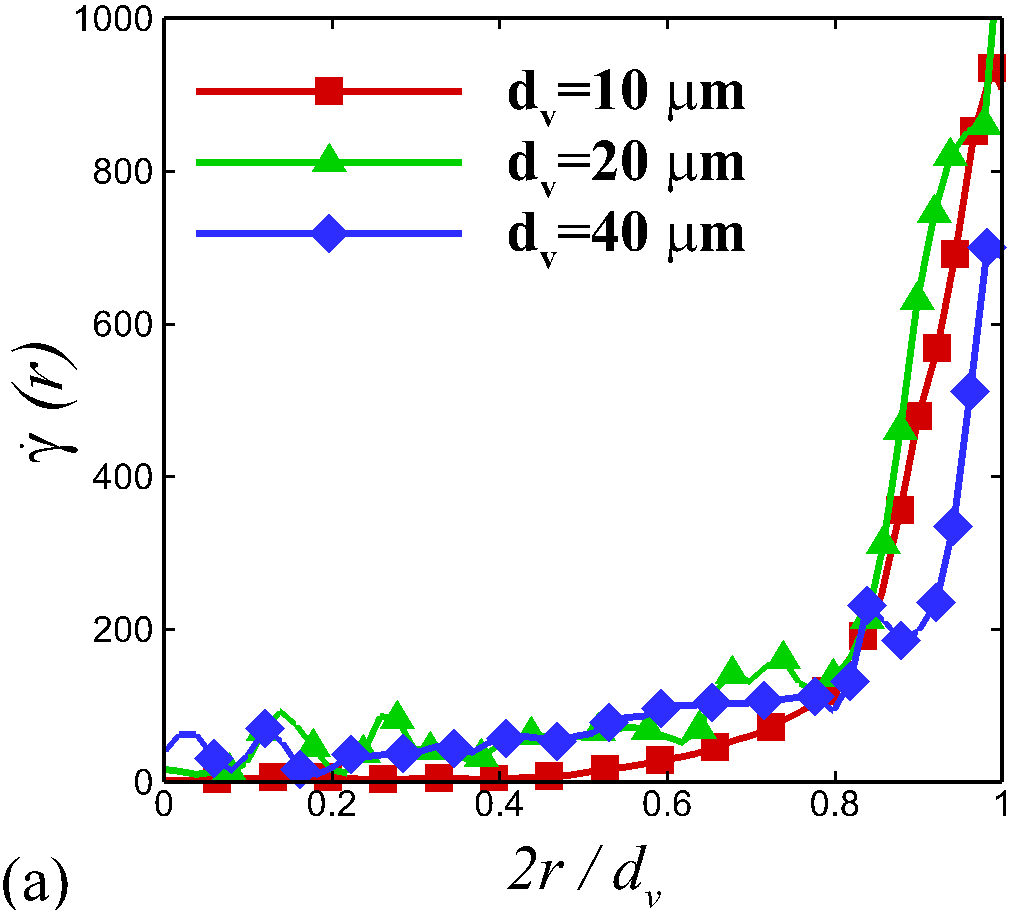}\  \  \  \  \  \  
 \includegraphics[width=0.47\linewidth]{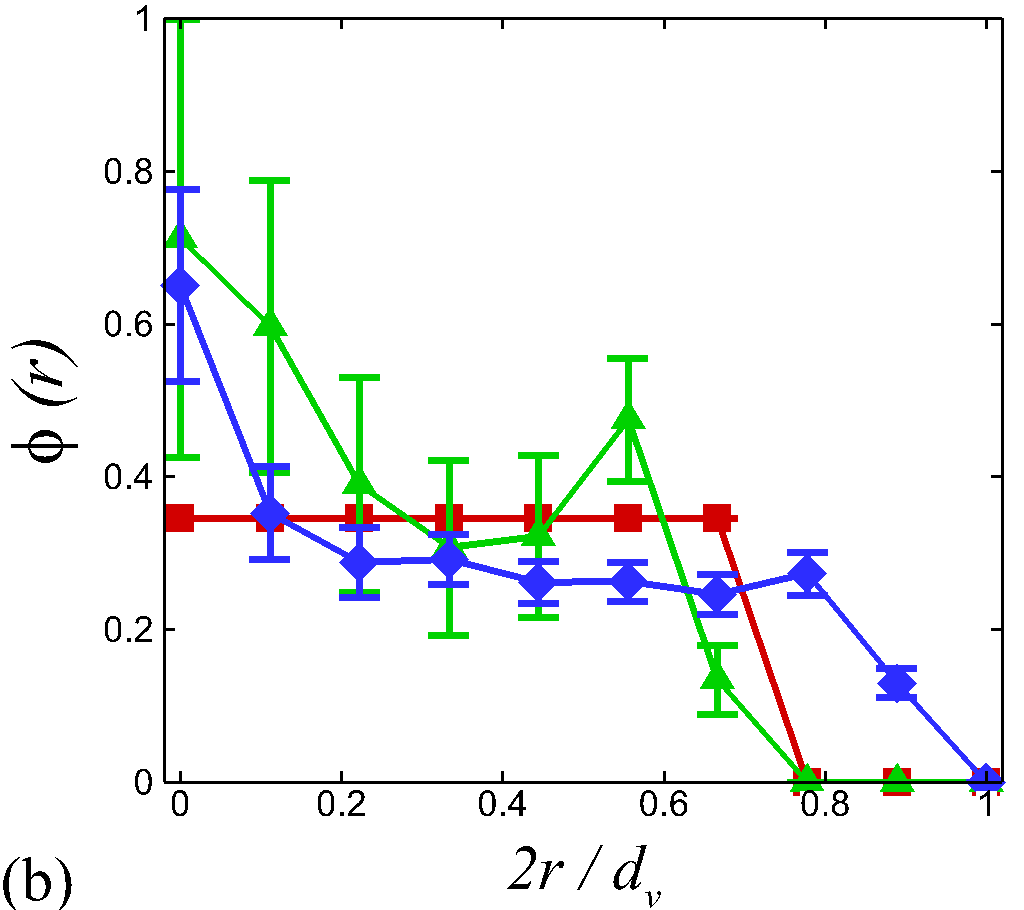}
 \includegraphics[width=0.47\linewidth]{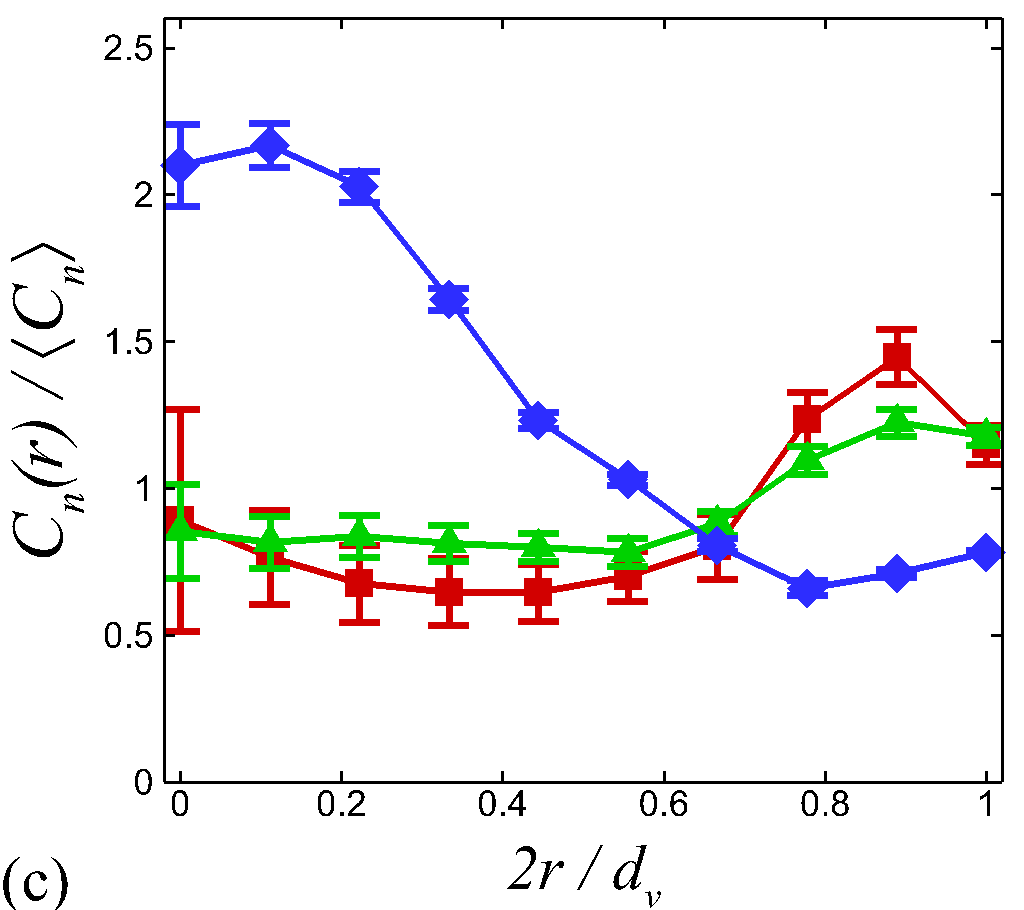}\  \  \  \  \  \  
 \includegraphics[width=0.47\linewidth]{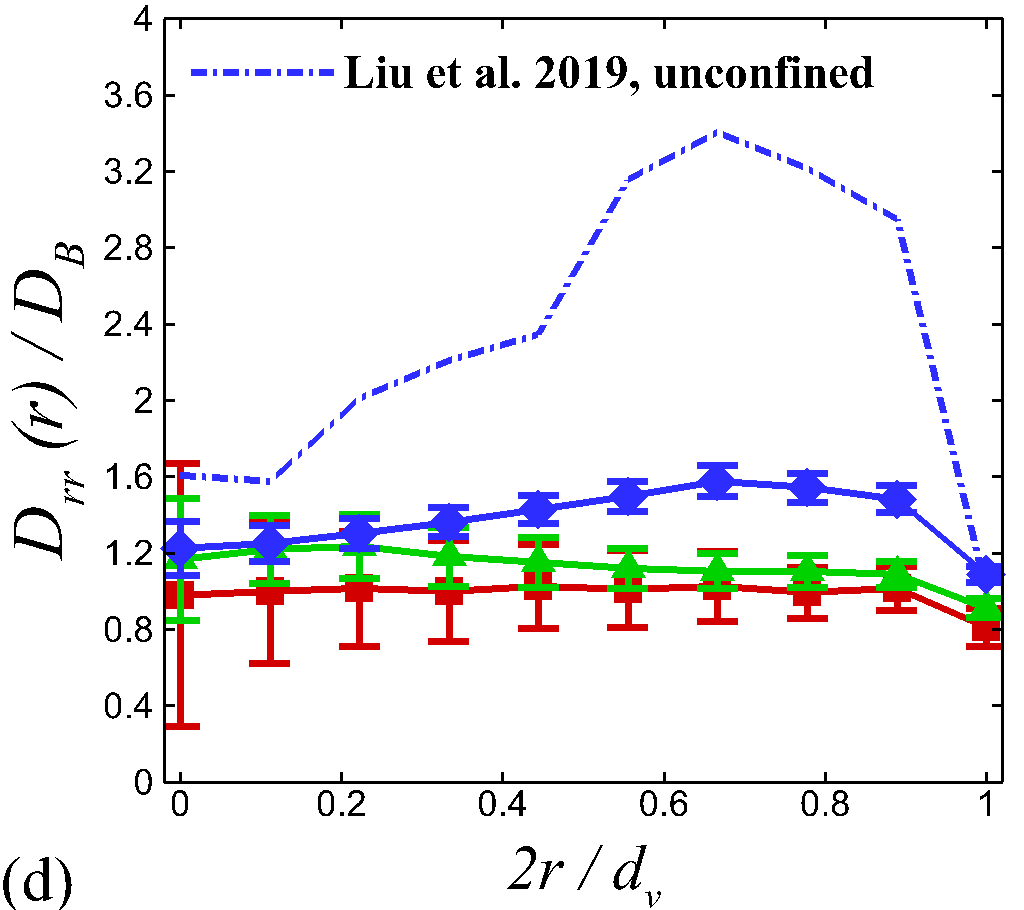}
\caption{\footnotesize Radial distribution of (a) shear rate, (b) hematocrit, (c) NP equilibrium distribution and (d) NP dispersion rate for various confinement ratios at $\dot{\gamma}_w=1000\ s^{-1}$, $\phi=0.2$ and $d_p=100\ nm$. The radial diffusivity based on the empirical correlation of NP diffusion tensor \citep{Liu2019} in a unconfined simple shear flow is plotted in (d) for comparison, where the calculation adopts the hemorheological parameters evaluated for the $d_v$=40 $\mu m$ case. Error bars denote the standard deviation.}
\label{fig:conf2}
\end{figure}

Fig~\ref{fig:conf1} presents the simulation snapshots of NP and RBC equilibrium distribution in microvessels under various confinement conditions. Qualitatively, the RBC dynamic mode changes from tank-treading/tumbling dominant to parachuting dominant \citep{sui2008,Reasor2012,tomaiuolo2012}, as the vessel diameter decreases from 40 to 10 $\mu m$. Such an increase of confinement does not alter the radial distribution of shear rate significantly but does change the radial distribution of local hematocrit to a large extent, as shown in Fig~\ref{fig:conf2}a and \ref{fig:conf2}b. Specifically, for the case with $d_v$=40 $\mu m$, the RBC-laden region shows a relatively uniform distribution except near the axis of the tube where the shear rate is close to zero. Consequently, the NP concentration, $C_n(r)$, at 0$<$$2r/d_v$$<$0.2 appears twice the bulk average NP concentration, $\langle C_n\rangle$, while $C_n(r)$ near the wall exbhit slightly lower values than $\langle C_n\rangle$. As the vessel diameter decreases to 20 $\mu m$, the dimensionless CFL thickness $\delta_{CFL}/d_v$ increases from $\sim$0.2 to $\sim$0.4, i.e., the RBC-laden region becomes relatively more focused. Moreover, the local hematocrits get intensified especially at the inner boundary of the CFL and at the tube axis. These hemorheological changes substantially affect the equilibrium radial distribution of $C_n(r)/\langle C_n\rangle$. As a result, the location of peak NP concentration shifts towards the CFL region, as shown in Fig~\ref{fig:conf2}c. Further confining the system to $d_v$=10 $\mu m$, allowing only one train of RBCs parachuting through the vessel, appears to slightly enhance the peak concentration of NP at the CFL region while decreasing the NP concentration at the RBC-laden region. Previous microfluidic experiments by~\citet{nott2011suspension} also show the enhancement of the number percentage of particles adhered to the wall when the width of a $\sim$40 $\mu m$ channel is reduced by half.

The distributions of the NP radial concentration can be better understood by evaluating the NP radial diffusivity, as depicted in Fig~\ref{fig:conf2}d. The NP diffusivity in the velocity-gradient direction based on the unconfined linearly sheared blood flow~\citep{Liu2019} is also plotted for comparison using shear rates and hematocrits of the $d_v$=40 $\mu m$ case. In general, increasing the confinement reduces the magnitude of NP radial diffusivity, where the unconfined case shows up to two folds the radial diffusivity, $D_{rr}(r)/D_B$, of the $d_v$=40 $\mu m$ case. Besides changing the magnitude of $D_{rr}(r)/D_B$, adjusting confinement ratio also alters the radial distribution of $D_{rr}(r)/D_B$. For the $d_v$=40 $\mu m$ case, the NP radial diffusivity shows high values near the CFL inner boundary and low value in the RBC-laden region, which is similar to the $D_{rr}(r)/D_B$ distribution in the unconfined case. This distribution of $D_{rr}(r)/D_B$ seems to be the cause of the low concentration of NPs near the CFL region and the high concentration at the RBC-core region. The increase of confinement (decrease of vessel diameter to 20 or 10 $\mu m$) renders the radial location of high $D_{rr}(r)/D_B$ to move towards the RBC-laden region, which appears to be responsible for the shift of the high NP concentration region towards the CFL, as observed in the high confinement cases ($d_v$=10 and 20 $\mu m$). 

Overall, the increase of confinement ratio enhances the NP near-wall concentration by inhibiting the NP diffusion near the wall. This however does not warrant the margination of NPs, given no excessive NP concentration ($C_n(r)<1.5\langle C_n\rangle$) is observed near the wall as the vessel confinement increases to capillary scale. It is noted that when vessel size decreases to capillary scale, retention of microscale particles in the RBC-induced recirculation is reported~\citep{takeishi2017capture}, which however is not observed in the current study with NPs under the studied hemorheological conditions.

\subsection{Dependence on hematocrit}\label{sec:ht}
Changing hematocrit significantly modifies the apparent viscosity of blood~\citep{FedosovPNAS2011,ReasorJFM2013} and could drastically influence the RBC-enhanced shear-induced diffusivity of NPs in sheared blood flow~\citep{Liu2019}. To understand how the variation of systemic hematocrit in microvessels changes the local hemorheology and hence the NP radial distribution, we investigate the hematocrit dependence of the NP radial dispersion behavior under various systemic hematocrits in the range of $\phi$=0$\sim$0.3. For the cases considered here, we select a fixed vessel diameter of $d_v$=20 $\mu m$ and a NP size of $d_p$=100 $nm$. The wall shear rate is set to $\dot{\gamma}_w$= 1000 $s^{-1}$. The number of NPs are set to $N$=1000.
\begin{figure}[h]
	\centering
	\includegraphics[width=0.7\linewidth]{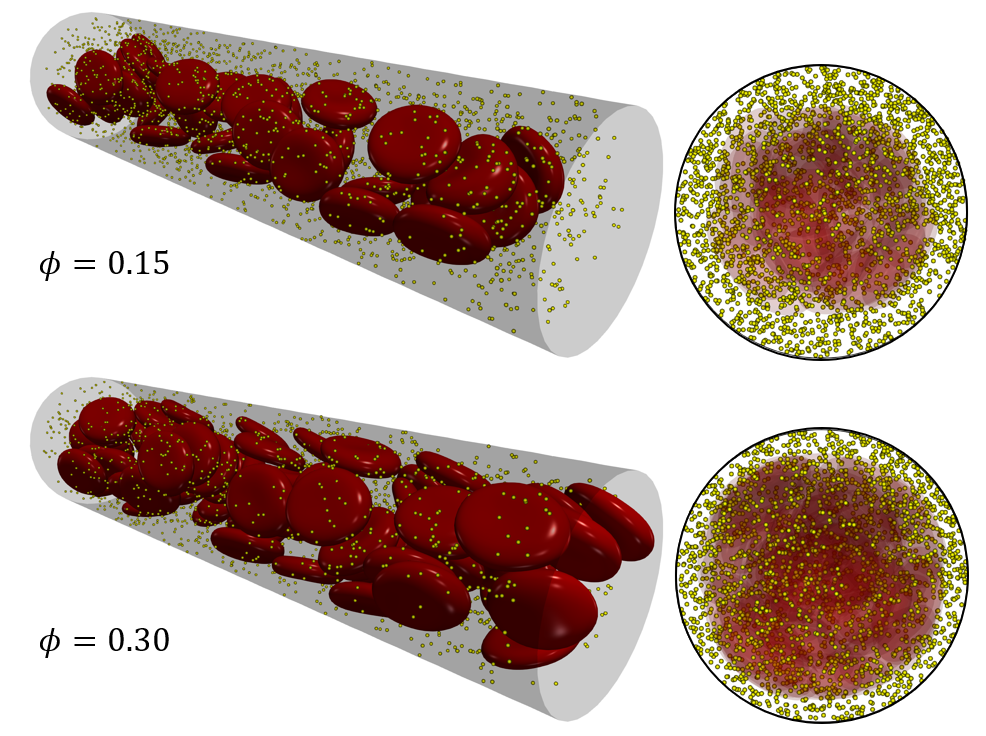}
\caption{\footnotesize NP and RBC distribution at equilibrium in a 20 $\mu m$ microvessel with $\phi=0.15$ (top) or $\phi=0.30$ (bottom) at $\dot{\gamma}_w=1000\ s^{-1}$ and $d_p=100\ nm$. Left column shows the isometric view of the tubular blood flow; right column shows the end view of the microvessels.}
	\label{fig:ht1}
\end{figure}

\begin{figure}[h]
\centering
   \includegraphics[width=0.47\linewidth]{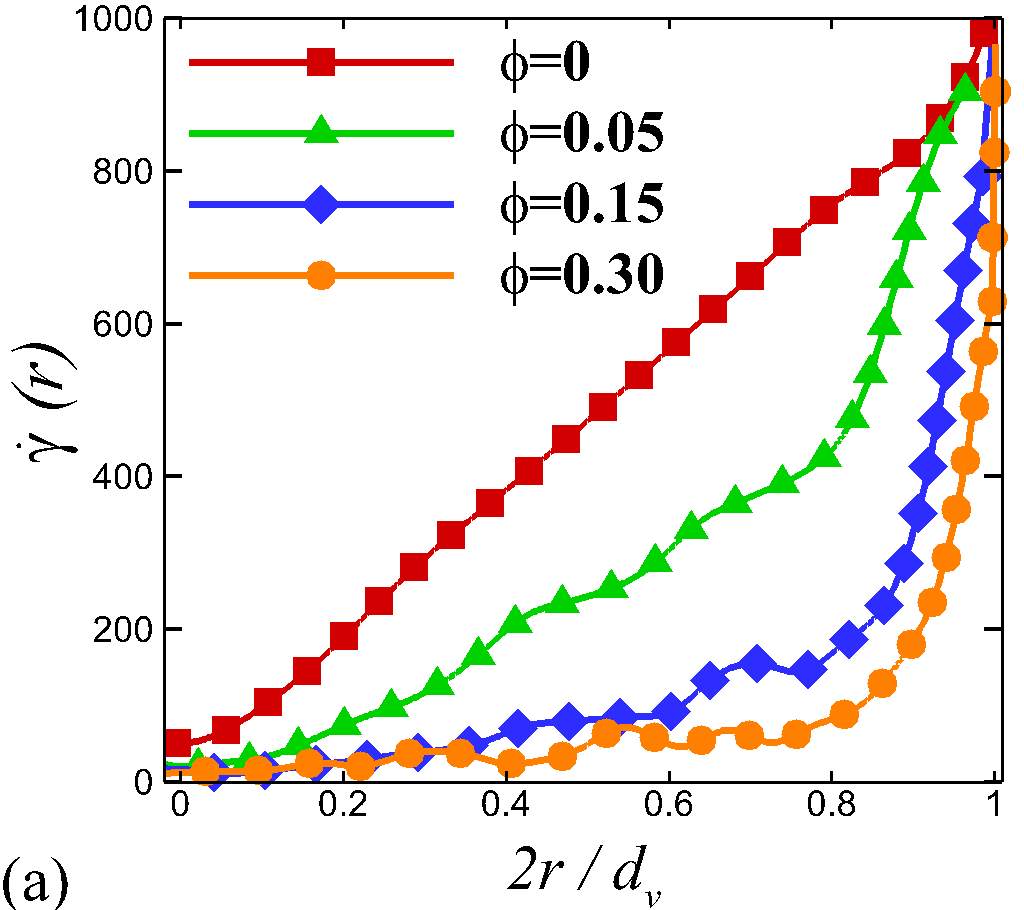}\  \  \  \  \  \  
   \includegraphics[width=0.47\linewidth]{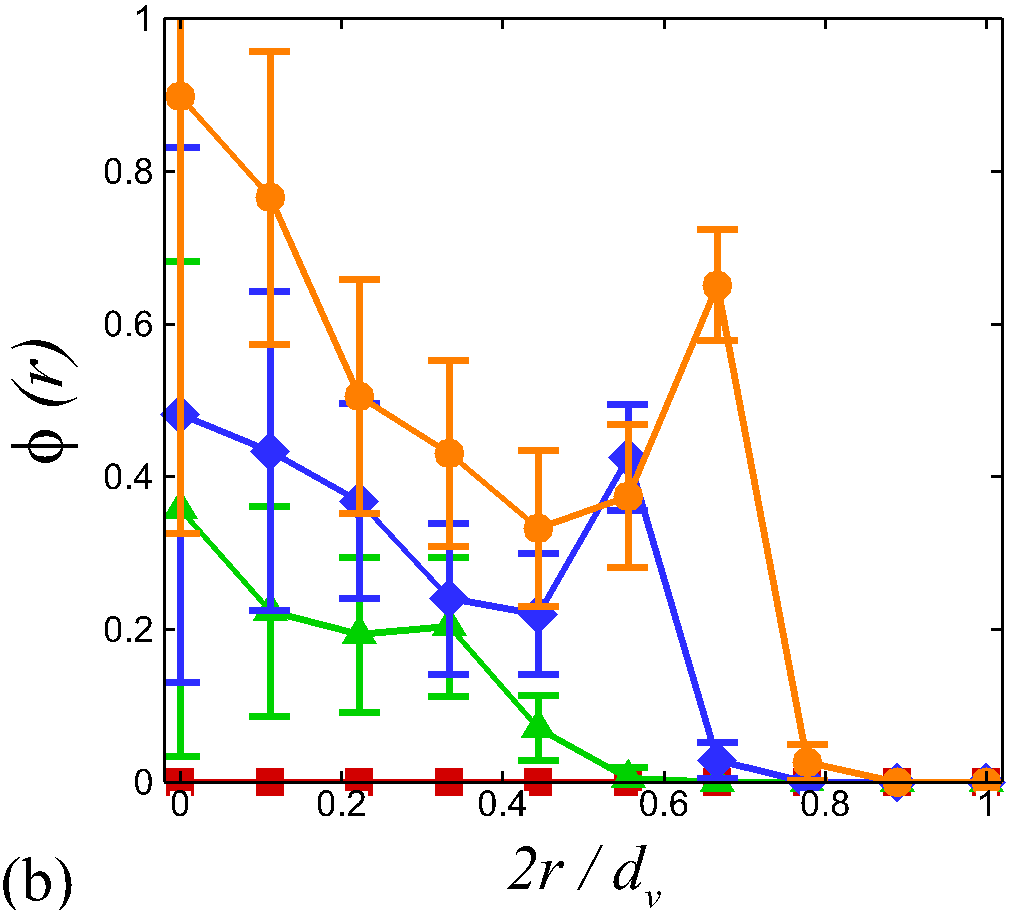}
   \includegraphics[width=0.47\linewidth]{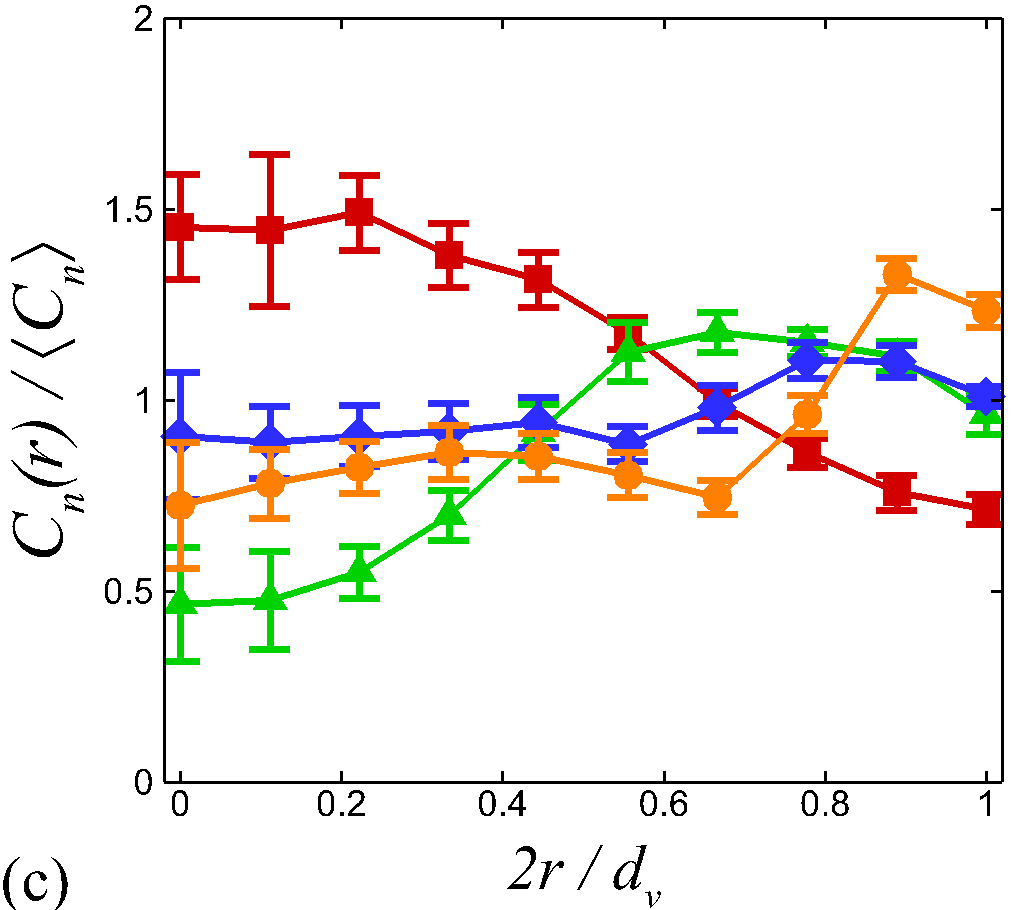}\  \  \  \  \  \  
   \includegraphics[width=0.47\linewidth]{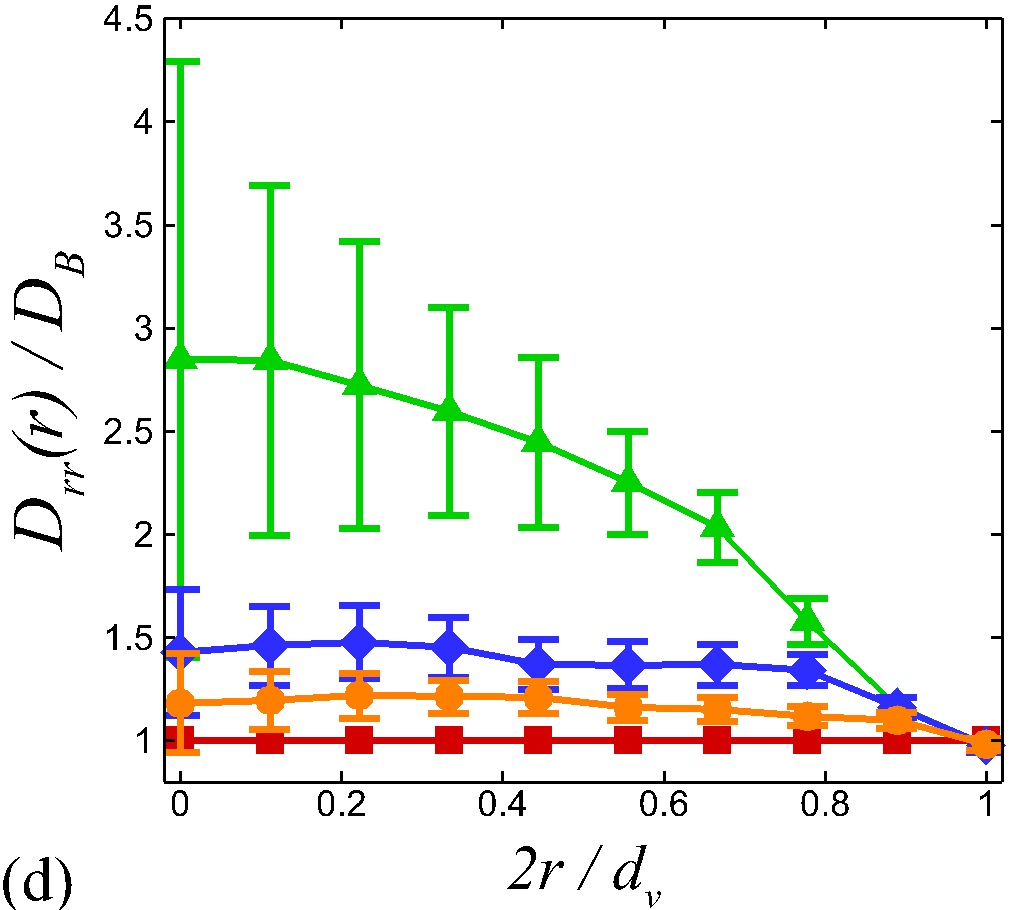}
\caption{\footnotesize Radial distribution of (a) shear rate, (b) hematocrit, (c) NP equilibrium distribution and (d) NP dispersion rate for various hematocrits at $\dot{\gamma}_w=1000\ s^{-1}$, $d_v=20\ \mu m$ and $d_p=100\ nm$. Error bars denote the standard deviation.}
\label{fig:ht2}
\end{figure}

Fig~\ref{fig:ht1} plots two snapshots of NP-RBC distribution in a 20 $\mu m$ vessel, where the high hematocrit case ($\phi$=0.3) exhbits a thinner CFL compared to the low hematocrit case ($\phi$=0.15) as expected. Fig~\ref{fig:ht2}a and ~\ref{fig:ht2}b present quantitative analysis of the hemorheological response to the change of systemic hematocrit, where the increase of systemic hematocrit affects the radial distribution of both the local shear rate, $\dot{\gamma}(r)$, and the local hematocrit, $\phi(r)$, which are two competing drivers for the particle cross-stream migration. On one hand, it alters the flow structure from a Poissuelle-type flow towards a plug-type flow; as a result, the local shear rate in the RBC-laden region decreases, which drives the Brownian particles towards the tube axis\citep{coupier2008}. One the other hand, it increases the local hematocrit in the RBC-laden region, which drives the particles to migrate to the wall.

The adjustment of the two competing effects lead to certain variation of the NP radial distribution, as presented in Fig~\ref{fig:ht2}c. At $\phi$=0, the NP dispersion is purely driven by the Brownian diffusivity and the shear-gradient driven dispersion. The former is isotropic, while the latter tends to drive the particle towards low shear region~\citep{nott2011suspension}. As a result, the NP distribution shows a high NP concentration in the core and a low concentration near the wall. Increasing the systemic hematocrit generally alters the NP distribution such that the high NP concentration region shifts to the CFL. Interestingly, a slight increase of $\phi$ from 0 to 0.05 appears to be enough to shift this paradigm of NP distribution, leading to about 3-fold decrease of NP concentration at the core and $\sim$1.5 folds increase of NP near wall concentration. Further increasing $\phi$ slightly increases the near wall NP concentration but also increases the NP concentration at the RBC-laden region. Correspondingly, in Fig~\ref{fig:ht2}d, the $D_{rr}(r)$ value (especially in the RBC-laden region) shows a non-monotonic change with respect to $\phi$, which first increases by up to 3 folds as $\phi$ rises to 0.05 and gradually gets inhibited to be close to the theoretical Brownian diffusivity as $\phi$ further increases to 0.3. The inhibition of NP radial diffusiviy at high hematocrit can be explained by the excessive local $\phi(r)$ and low $\dot{\gamma}(r)$, as shown in Fig~\ref{fig:ht2}a and~\ref{fig:ht2}b.

Therefore, low systemic hematocrits appear to be optimal to enhance the NP near-wall concentration in microvessels, owing to the relatively high local shear rates and moderate local hematocrits that does not inhibits the NP dispersion in the tubular core. Nevertheless, changing hematocrit does not lead to the margination of NP.

\subsection{Dependence on particle size}\label{sec:size}
\textcolor{black}{So far we have focused on the long-time dispersion behavior of NPs in microvessels under various confinement ratios and hematocrit conditions. In these cases, particles do not show margination behavior. Instead, a non-uniform radial distribution of particles is observed with the particle concentration near the wall being less than 1.5 times its bulk average concentration. Besides, the near-wall concentration of NP is dynamically conserved at equilibrium, accompanied by the cross-migration of NPs between the CFL and the RBC-laden region due to severe Brownian effect\citep{Liu2018a}.} 

In this section, we consider the size-dependent dispersion behavior of nano-to-microscale particles in microvessels. Particles with sizes ranging from $d_p$=10$\sim$2500 $nm$ are considered, covering particles ranging from nanoscale biomolecules such as vWFs in globular conformation to microscale cells such as platelets. The vessel diameter is fixed to $d_v$= 20 $\mu m$. The wall shear rate is set to 1000 $s^{-1}$. The systemic hematocrit is kept at $\phi$=0.2. \textcolor{black}{To maintain the volumetric concentration of the particle phase in the dilution limit (i.e., $\ll 1\%$), the number of large particles considered in the system is reduced accordingly but kept above 50 to ensure statistical significance as consistent with our previous margination study~\citep{ReasorABE2013}.}
\begin{figure}[h]
\centering
\includegraphics[width=0.7\linewidth]{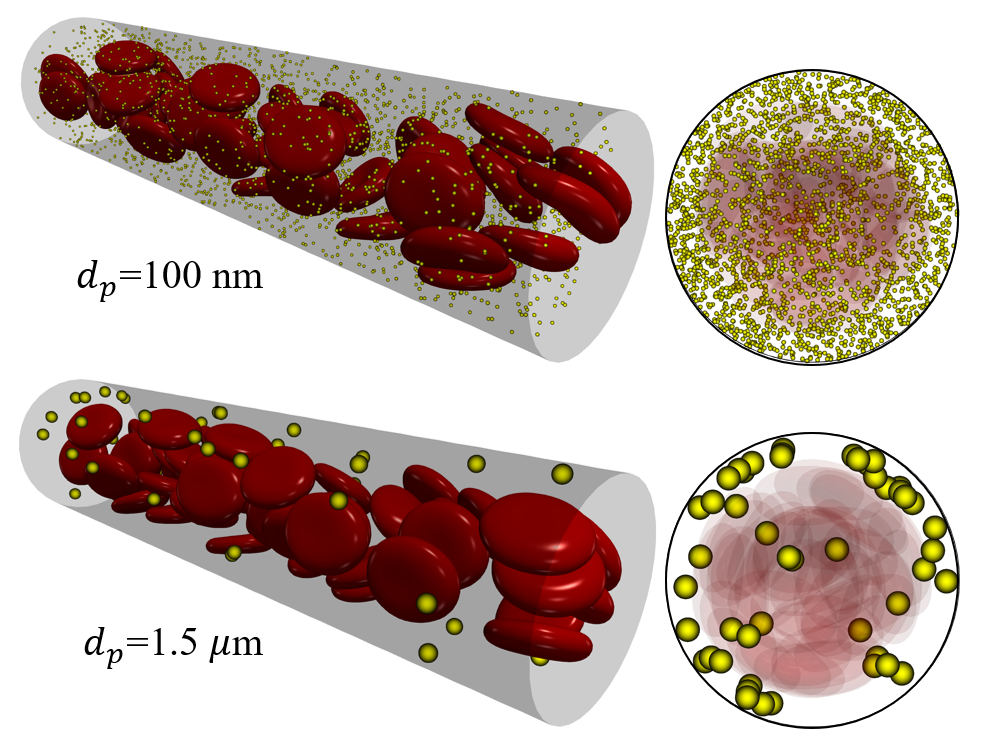}
\caption{\footnotesize Particle and RBC distribution at equilibrium with particles size being nanoscale (top) or microscale (bottom) at $\phi=0.2$, $d_v=20\ \mu m$ and $\dot{\gamma}_w=1000\ s^{-1}$. Left column shows the isometric view of the tubular blood flow; right column shows the end view of the microvessels.}
\label{fig:size1}
\end{figure}
\begin{figure}[h]
\centering
 \includegraphics[width=0.47\linewidth]{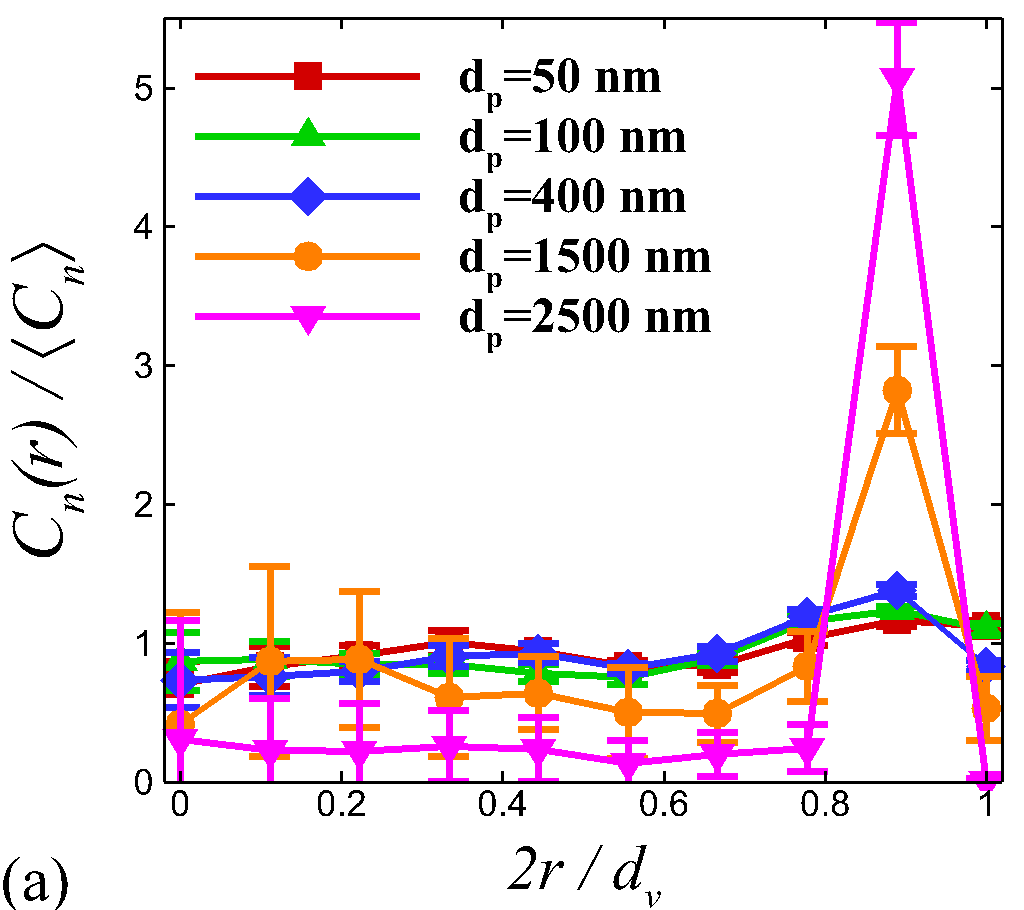}\  \  \  \  \  \  
 \includegraphics[width=0.47\linewidth]{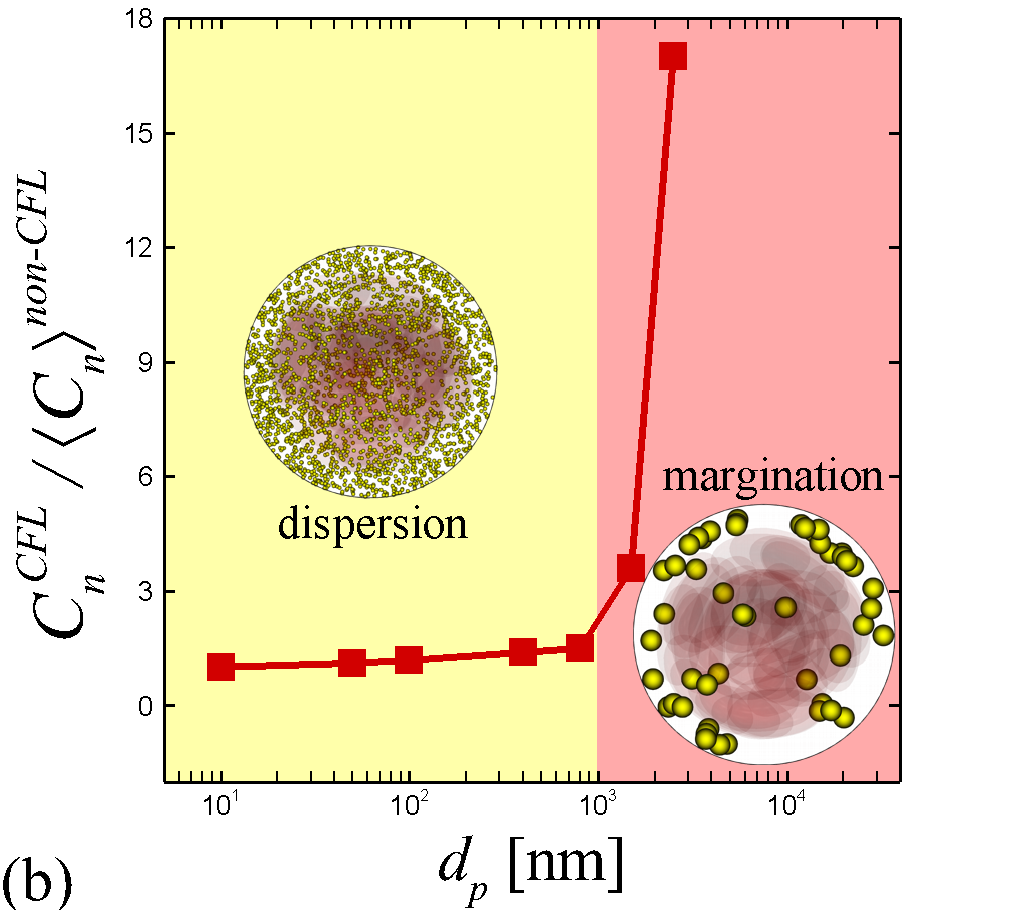}
 \includegraphics[width=0.47\linewidth]{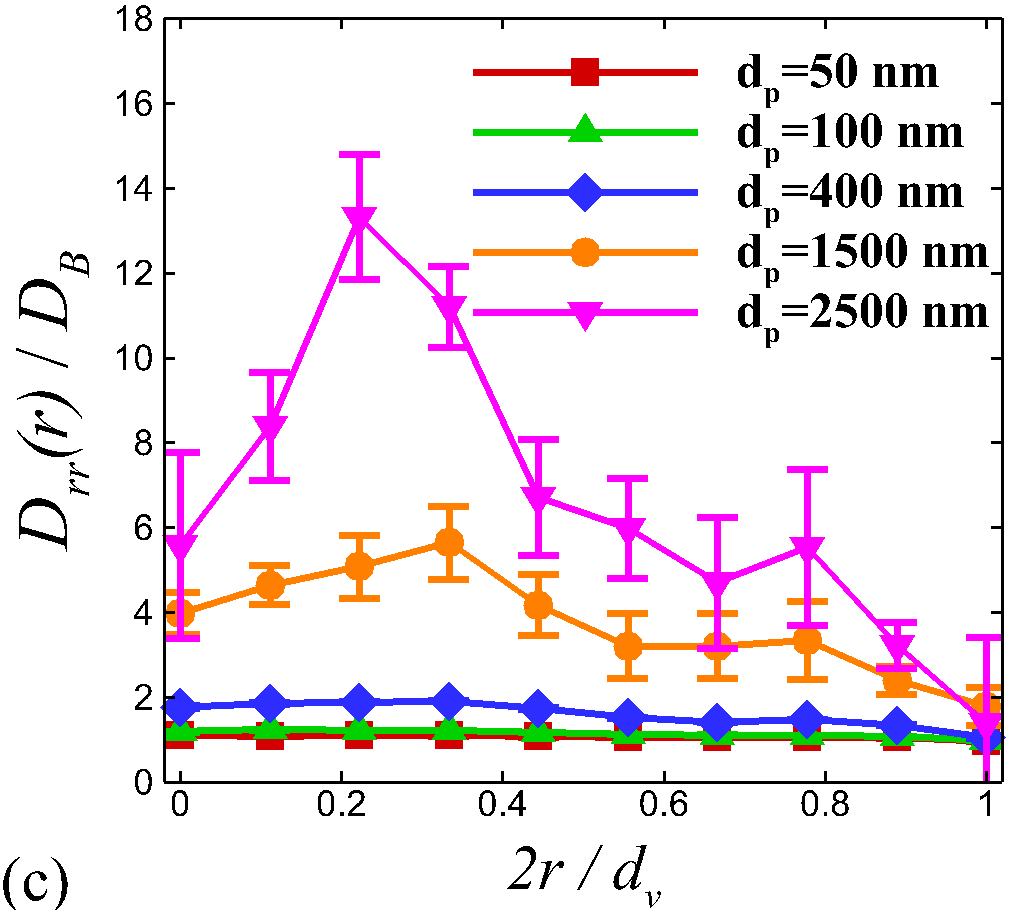}\  \  \  \  \  \  
 \includegraphics[width=0.47\linewidth]{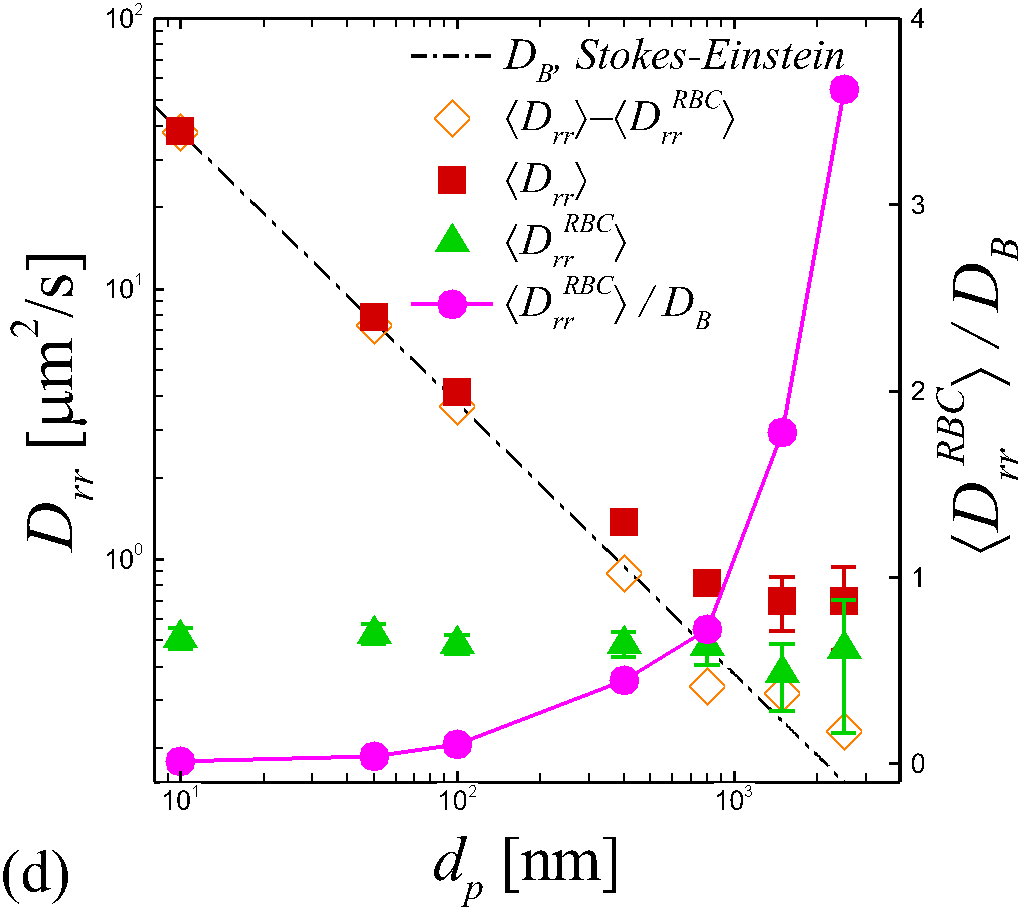}
\caption{\footnotesize \textcolor{black}{(a) The radial distribution of particle number concentration normalized by the bulk average number concentration of the particles for different particle sizes at $\phi=0.2$, $d_v=20\ \mu m$ and $\dot{\gamma}_w=1000\ s^{-1}$. (b) The particle number concentration in the CFL normalized by 
%the bulk average particle concentration, $C_n^{CFL}/\langle C_n\rangle$,
that in the RBC-laden region, $C_n^{CFL}/\langle C_n\rangle^{non-CFL}$, 
plotted against particle size. The yellow area shows the dispersion (no margination) regime; the pink area show the margination regime. (c) The radial distribution of particle radial diffusivity normalized by the Brownian diffusivity for various particle sizes. (d) The ensemble-averaged particle radial diffusivity plotted against particle sizes; the diffusivity ratio, $\langle D_{rr}^{RBC}\rangle$, is also plotted with the vertical axis on the right. Error bars denote the standard deviation.}}
\label{fig:size2}
\end{figure}

\textcolor{black}{Fig~\ref{fig:size1} presents the equilibrium distributions of RBCs and particles. Qualitatively, NPs show non-uniformly dispersed distribution across the vessel, where NPs at any radial position can disperse to a random radial location given enough time owing to severe Brownian effect. These features are qualitatively different from the margination behavior of microscale particles, where ecessive concentration and retention of particles in the CFL can be observed\citep{ReasorABE2013,MM2016}.} Fig~\ref{fig:size2}a further depicts the radial distribution of $C_n(r)/\langle C_n\rangle$ for various particle sizes. As the particle size increases above 1 $\mu m$, the CFL region exhibits a prominently high particle concentration. Specifically, for particles with a diameter $d_v$=2.5 $\mu m$, a five times bulk average particle concentration, $C_n(r)\approx5\langle C_n\rangle$, can be observed at the CFL region. These observations are consistent with particle margination study using a microfluidic perfusion system by \citet{namdee2013margination}, where the number percentage of particles adhered to the wall gets increased by 5 to 7 times as the particle size changes from nanoscale to microscale. \textcolor{black}{The change of the particle number concentration at the CFL, $C_n^{CFL}$, versus that at the RBC-laden region, $\langle C_n\rangle^{non-CFL}$, as a function of particle size is further plotted in Fig~\ref{fig:size2}b. When the particle size is below 1 $\mu m$, the $C_n^{CFL}/\langle C_n\rangle^{non-CFL}$ value shows weak dependence on the particle size with only a slight increase from 1.0 to about 1.5 as $d_p$ changing from 10 nm to 1000 nm. As particle size exceeds 1 $\mu m$, the $C_n^{CFL}/\langle C_n\rangle^{non-CFL}$ value increases abruptly (up to $\sim$17) and margination occurs. Particle size $d_p$=1 $\mu m$ seems to be a critical watershed that divides the $dispersion$ state and the $margination$ state, as denoted in Fig~\ref{fig:size2}b.}

To shed light on the size-dependent dispersion behavior of particles in tubular blood flows, the distribution of particle radial diffusivity, $D_{rr}(r)/D_B$, is plotted in Fig~\ref{fig:size2}c. For NPs, the $D_{rr}(r)/D_B$ distribution tends to be uniform due to the dominance of isotropic Brownian diffusivity. As the particle size increases above one micronmetre, the RBC-laden region shows prominent enhancement of $D_{rr}(r)/D_B$ compared to the CFL. Moreover, both at the CFL edge ($2r/d_v\sim 0.8$) and close to the tube axis ($2r/d_v\sim 0.2$), the magnitude of $D_{rr}(r)/D_B$ peaks and the inner peak is more pronounced than the peak close to the CFL. For the 2500 nm particles, the inner peak shows more than ten times $D_{rr}(r)/D_B$ values of that at the CFL. The radial distribution of $D_{rr}(r)/D_B$ seems to be inversely correlated to the radial distribution of $C_n(r)/\langle C_n\rangle$ in terms of the radial location, suggesting that the margination of microscale particles is probably due to the large magnitude difference in $D_{rr}(r)/D_B$ between the RBC-laden region and the CFL region.

In Fig~\ref{fig:size2}d, we plot the ensemble-averaged radial diffusivity, $\langle D_{rr}\rangle$, as a function of the particle size. The ensemble average is performed among all particles located at various radial locations at equilibrium state. For small $d_p$, the $\langle D_{rr}\rangle$ value asymptotically matches the Stokes-Einstein relation due to the dominance of Brownian diffusion. Increasing the particle size decreases the effect of thermal fluctuation and leads to the deviation of $\langle D_{rr}\rangle$ from $D_B$. The value of $\langle D_{rr}\rangle$ eventually plateaus at the microscale size regime, where the RESID is dominant over BD. The bulk ensemble-averaged RESID, $\langle D_{rr}^{RBC}\rangle$, seems to be weakly dependent on the particle size, similar to the particle diffusivity observed in a unbounded sheared blood flow~\cite{Liu2019}. Subtracting the $\langle D_{rr}\rangle$ with $\langle D_{rr}^{RBC}\rangle$ shows an overlap of the dataset $\langle D_{rr}\rangle$-$\langle D_{rr}^{RBC}\rangle$ with the theoretical Brownian diffusivty, confirming the RESID is linearly superimposed with the Browanian diffusivity~\cite{Liu2018a}. \textcolor{black}{More interestingly, the increase of $\langle D_{rr}^{RBC}\rangle/D_B$ is strongly correlated with the increase of $C_n^{CFL}/\langle C_n\rangle^{non-CFL}$, as shown in Fig~\ref{fig:size2}b and \ref{fig:size2}d. This strong linkage between the dispersion-to-margination transition and the particle-size relevant change of RESID-to-Brownian diffusivity ratio is consistent with the margination characterization based on a kinetic theory-based analysis \citep{Graham2016PRF}, where the margination of microscale particles is shown to be weakened as strong Brownian effect comes into play.} 

%%%%%%%%%%%%%%%%%%%%%%%%%%%%%%%%%%%%%%%%%%%%%%%%%%%%%%%%%%%%%%%%%%%%%%%%%%%%%%
%%%%%%%                           Conclusions                           %%%%%%
%%%%%%%%%%%%%%%%%%%%%%%%%%%%%%%%%%%%%%%%%%%%%%%%%%%%%%%%%%%%%%%%%%%%%%%%%%%%%%

\section{Conclusions}\label{sec:conclusion}
Using a three-dimensional multiscale complex blood flow solver~\cite{Reasor2012,Liu2018a,Liu2018b}, we have interrogated the long-time dispersive characteristics of rigid spherical particles with sizes across nano-to-micrometers in blood flow through microvessels. The role of the confinement ratio and the systemic hematocrit in altering the nanoparticle radial dispersion is quantitatively analyzed in terms of the radial distribution of particle concentration and particle radial diffusion rate. The effect of changing particle size on the alteration of the particle dispersive characteristics is highlighted.

\textcolor{black}{In the range of parameters considered here, it is found that nanoscale particles do not marginate under various confinement effects or hematocrit levels in the same way as microscale particles do, but rather show a non-uniform radial distribution across the vessel.} Increasing the confinement effect by decreasing the vessel diameter hinders the particle radial diffusivity but also enhances the equilibrium concentration of nanoscale particles in the cell free layer. Low hematocrit level ($\phi\sim5\%$) in the microvessel appears to be optimal to the radial dispersion of nanoscale particles, leading to high radial diffusion rate and near-wall concentrations being higher than the average concentration. High hematocrits ($\phi$=30$\%$) slightly increases the near-wall concentration but meantime inhibits the dispersion of nanoscale particles in the RBC-laden region.

\textcolor{black}{Microscale particles exhibit pronounced margination behavior, where at equilibrium the microscale particles get concentrated in the cell-free layer at up to 5 times the particle average concentration in the bulk (or more than 10 times the particle concentration in the RBC-laden region).} The margination propensity seems to be enhanced with the particle size. For microscale particles, the RBC-enhanced shear-induced diffusivity is dominant over the Brownian diffusivity, where the RBC-laden area shows more than 10 times higher diffusivity compared to that in the RBC-free layer. The particle-size induced alteration of particle radial diffusivity in both distribution and magnitude gives rise to margination of microscale particles in confined tubular blood flows.

\begin{acknowledgments}
The authors acknowledge the partial support from Sandia National Laboratories under grant 2506X36 and the compuatational resource provided by National Science Foundation under grant TG-CT100012. Sandia National Laboratories is a multimission laboratory managed and operated by National Technology and Engineering Solutions of Sandia LLC, a wholly owned subsidiary of Honeywell International Inc. for the U.S. Department of Energy’s National Nuclear Security Administration under contract DE-NA0003525. This paper describes objective technical results and analysis. Any subjective views or opinions that might be expressed in the paper do not necessarily represent the views of the U.S. Department of Energy or the United States Government.
\end{acknowledgments}

\bibliography{biblio}% Produces the bibliography via BibTeX.

\end{document}